\documentclass[10pt]{article}
\usepackage{fullpage}
\usepackage{amsmath}
\usepackage{amssymb}
\usepackage[dvips]{epsfig}
\usepackage{color}

\def\##1{\underline{#1}}
\def\=#1{\underline{\underline{#1}}}

\def\+#1{\underline{\bf #1}}
\def\*#1{\underline{\underline{\bf #1}}}

\def\r#1{(\ref{#1})}
\def\l#1{\label{#1}}
\def\c#1{\cite{#1}}

\def\le{\left(}
\def\ri{\right)}
\def\les{\left[}
\def\ris{\right]}
\def\lec{\left\{}
\def\ric{\right\}}

\def\.{\cdot}

\def\epso{\epsilon_{\scriptscriptstyle 0}}

\def\muo{\mu_{\scriptscriptstyle 0}}

\def\ko{k_{\scriptscriptstyle 0}}
\def\co{c_{\scriptscriptstyle 0}}
\def\Eo{\#E_{\, \scriptscriptstyle 0}}
\def\Ho{\#H_{\, \scriptscriptstyle 0}}

\begin{document}

\begin{center}

{\bf {\LARGE Ray trajectories for Alcubierre spacetime}}

\vspace{10mm} \large

 Tom H. Anderson\footnote{E--mail: T.H.Anderson@sms.ed.ac.uk}\\
{\em School of Mathematics and
   Maxwell Institute for Mathematical Sciences\\
University of Edinburgh, Edinburgh EH9 3JZ, UK}\\
 \vspace{3mm}
 Tom G. Mackay\footnote{E--mail: T.Mackay@ed.ac.uk}\\
{\em School of Mathematics and
   Maxwell Institute for Mathematical Sciences\\
University of Edinburgh, Edinburgh EH9 3JZ, UK}\\
and\\
 {\em NanoMM~---~Nanoengineered Metamaterials Group\\ Department of Engineering Science and Mechanics\\
Pennsylvania State University, University Park, PA 16802--6812,
USA}\\
 \vspace{3mm}
 Akhlesh  Lakhtakia\footnote{E--mail: akhlesh@psu.edu}\\
 {\em NanoMM~---~Nanoengineered Metamaterials Group\\ Department of Engineering Science and Mechanics\\
Pennsylvania State University, University Park, PA 16802--6812, USA}

\end{center}

\vspace{4mm}

\normalsize

\begin{abstract}

The Alcubierre spacetime was simulated by means of a Tamm
medium which is asymptotically identical to vacuum and has
constitutive parameters which are  continuous functions of the
spatial coordinates. Accordingly, the Tamm medium is amenable to
physical realization as a nanostructured metamaterial. A
comprehensive characterization of ray trajectories in the Tamm
medium was undertaken, within the geometric-optics regime.
Propagation directions corresponding to evanescent waves  were
identified: these occur in the region of the Tamm medium which
corresponds to the \textit{warp bubble} of the Alcubierre spacetime, especially
for directions perpendicular to the velocity of the warp bubble  at high speeds of that
bubble. Ray trajectories are acutely sensitive to the
magnitude and direction of the warp bubble's velocity, but rather
less sensitive to the thickness of the transition zone between the
warp bubble and its background. In particular, for rays which
travel in the same direction as the warp bubble, the latter
acts as a focusing lens, most notably at high   speeds.

\end{abstract}

\vspace{5mm} \noindent  PACS numbers: 04.40.Nr, 41.20.-q, 42.15.-i

\vspace{5mm} \noindent  {\bf Keywords:} Alcubierre warp drive, Tamm
medium,  metamaterial, ray tracing

\section{Introduction}

Metamaterials provide opportunities to study general-relativistic
scenarios which would otherwise be either impractical or impossible to
explore \c{Smolyaninov_JOP}. This may be achieved by exploiting the
formal analogy that exists between light propagation in vacuum
subjected to a gravitational field and light propagation in a
certain nonhomogeneous bianisotropic medium, known as a Tamm medium
\c{Skrotskii,Plebanski,SS}. The practical creation of Tamm mediums is beginning to look like
 an increasingly realistic proposition, as rapid developments
are made in the science of nanostructured metamaterials. In recent
years, metamaterial-based simulations of  black holes
\c{Smolyaninov_NJP},
 cosmic
strings \c{Spinning_string}, de Sitter spacetime \c{Li_1,Li_2} and
wormholes \c{Greenleaf} have been proposed, for example.

As a useful \emph{Gedankenexperiment}, the spacetime of the
Alcubierre metric has generated considerable interest since its
introduction in 1994 \c{Alcubierre,Clark_CQG,Natario}. The spacetime
is characterized by a \textit{warp bubble} which moves with respect to an
asymptotically flat background  at a  speed $ v'_s \co $, where $\co$ is
the speed  of light in the absence of a gravitational field and $v'_s\geq0$. By effectively
contracting the spacetime ahead of the  warp bubble and expanding
the spacetime behind it, arbitrary relative speeds $v'_s$ may be achieved, in
principle. In contrast, within the local environment of the warp
bubble, its speed is necessarily subluminal; i.e., $v'_s < 1$.

Serious obstacles stand in the way of a physical realization
of the Alucbierre spacetime, stemming from  violations of various energy
constraints \c{Pfenning,Broeck}. Even the subluminal
 regime is associated with negative energy densities \c{Lobo_CQG}.
However, negative energy densities are not unprecedented in
astrophysics. For examples,  the emission of Hawking radiation by a
black hole is accompanied by a flow of negative energy \c{Hawking},
and the construction of wormholes relies on negative energy density
\c{Safonova}. Nor is negative energy density unprecedented within
the realm of metamaterials: propagation of (monochromatic) plane
waves with negative phase velocity \c{EJP}~---~which is intimately
related to the phenomenon of negative refraction \c{ML_PRB} that
certain metamaterials have been shown to support in experimental
observations \c{SSS}~---~also involves negative energy density
\c{Ziol}, at least in the absence of dissipation \c{Rupp}.

In this communication, we present a
  simulation of the
Alcubierre spacetime, in the form a Tamm medium which is physically
realizable,  in principle. Under the geometric-optics approximation,
a comprehensive characterization  of light ray trajectories through
the Tamm medium is provided.

As regards notation, $3$--vectors are underlined and unit $3$-vectors are
additionally distinguished by a caret, whereas   3$\times$3
dyadics \c{Chen} are double underlined. The identity dyadic is written as
$\=I$. The scalar constants $\epso$ and $\muo$ denote the
permittivity and permeability of vacuum in the absence of a
gravitational field, and $\co = 1 / \sqrt{\epso \muo}$.

\section{Tamm medium for Alcubierre spacetime}

The Alcubierre spacetime is specified by the line element \c{Alcubierre}
\begin{equation} \l{le}
ds^2 = - \co^2 dt'^2 + \les dx' - v'_s f(r'_s) \co dt' \ris^2 +
dy'^2 + dz'^2,
\end{equation}
for the case where the  warp bubble travels at relative velocity $\#v'_s =
v'_s \, \hat{\#x}'$. Herein the scalar function
\begin{equation} \l{f_def}
f (r'_s) = \frac{\tanh \les \sigma \le r'_s + R \ri \ris -  \tanh
\les \sigma \le r'_s - R \ri \ris }{ 2 \tanh \le \sigma R \ri }
\end{equation}
is unit--valued at the origin and decays uniformly to zero as $r'_s
\to  \infty$, where $x'_s (t') = v'_s \co t' $ and
 the time-dependent translated displacement
 \begin{equation}
r'_s (t' ) = \sqrt{ \les x' - x'_s (t') \ris^2 + y'^2 + z'^2}\,.
\end{equation}
The scalar parameter $R > 0$ is a measure of the warp bubble radius
whereas $\sigma > 0$ is a measure of the thickness of the transition
zone between the warp bubble and its background. For $\sigma \gg R$,
the function $f$ has an approximately top-hat profile which
propagates along the positive $x'$ axis at relative speed $v'_s$.

In order to eliminate the time-dependency which enters via  $x'_s
(t')$, let us introduce the spacetime coordinates
 \c{Chen}
\begin{equation}
\left. \l{Lorentz}
\begin{array}{l}
t =  \displaystyle{ \nu \, \le \, t' - \frac{ x' v'_s}{\co} \,
 \ri}\\
x = \nu \le x' -  v'_s \co t' \ri \\
y = y' \\
z = z'
\end{array}
\right\},
\end{equation}
with the scalar quantity
\begin{eqnarray}
&&  \nu = \frac{1}{\sqrt{1 - \le v'_s \ri^2}}.
\end{eqnarray}
This change amounts to a Lorentz transformation \c{Chen}.
 The line element \r{le} may then be expressed as
\begin{eqnarray} \l{lep} ds^2 &=&  \lec \frac{ \les f(r_s) - 1 \ris^2 \le v'_s \ri^2 -1}{1 - \le v'_s \ri^2} \ric \co^2  dt^2
+ \frac{2 f(r_s)  v'_s \lec   \les f(r_s)  -1  \ris \le v'_s \ri^2
-1 \ric}{1 -
\le v'_s \ri^2} \,\co dt dx \nonumber \\
&& + \frac{1 + \le v'_s \ri^2 \lec  f(r_s)  \les  f(r_s)  \le v'_s
\ri^2 -2\ris -1 \ric}{1 - \le v'_s \ri^2} \, dx^2
 + dy^2 + dz^2,
\end{eqnarray}
wherein
\begin{equation}
r_s  = \sqrt{  \frac{x^2}{\nu^2} + y^2 + z^2}.
\end{equation}
is independent of $t$.

Per the noncovariant approach of Tamm \c{Skrotskii,Plebanski,SS},
electromagnetic fields in
 curved  spacetime  may be described by the constitutive
 relations
\begin{equation}
\label{CR2} \left.
\begin{array}{l}
\#D (\#r, t)  =  \epso \=\gamma_{\,\mbox{\tiny{}}}(\#r) \. \#E(\#r,
t) - \sqrt{\epso \muo}\,
\#\Gamma_{\,\mbox{\tiny{}}}(\#r) \times \#H (\#r, t) \vspace{8pt}\\
\#B (\#r, t)  = \sqrt{\epso \muo}\,  \#\Gamma_{\,\mbox{\tiny{}}}(\#r) \times \#E(\#r, t)  +  \muo \=\gamma_{\,\mbox{\tiny{}}}(\#r) \. \#H (\#r, t)\\
\end{array}\right\}
\end{equation}
of an equivalent medium in flat spacetime, using SI units, with $\#r=
x\, \hat{\#x}+y\, \hat{\#y}+z\, \hat{\#z}$.
  The components of the 3$\times$3
dyadic $\=\gamma (\#r)$ and the $3$-vector $\#\Gamma (\#r)$  are
defined in indicial notation as
\begin{equation}
\label{akh1} \left.\begin{array}{l} \gamma_{\ell m}
= \displaystyle{ \sqrt{ -g}  \, \frac{{g}^{\ell m}}{{g}_{00}}}\\[6pt]
\Gamma_m= \displaystyle{\frac{g_{0m}}{g_{00}}}
\end{array}\ric, \qquad (\ell, m \in \lec 1, 2, 3 \ric),
\end{equation}
where $g$ denotes the determinant of spacetime metric $g_{\alpha
\beta}$,  $(\alpha, \beta \in \lec 0, 1, 2, 3 \ric)$, prescribing
the curved spacetime. The fictitious Tamm medium represented by the
constitutive relations \r{CR2} is spatiotemporally
  local, and generally nonhomogeneous and  bianisotropic.
 The sign of the square root term in the
definition of $\gamma_{\ell m}$ is selected such that the metric for
vacuous Minkowskian spacetime is represented by the dyadic $\=\gamma
= \=I $.

For the case of  Alcubierre spacetime characterized by the line
element \r{lep}, the definitions \r{akh1} deliver
\begin{equation} \l{gamma_dyadic}
 \=\gamma (\#r) = \hat{\#x}\, \hat{\#x} +  \frac{1 - \le v'_s \ri^2}{1 - \les  1 - f(r_s) \ris^2 \le v'_s \ri^2
 } \le  \hat{\#y}\, \hat{\#y} + \hat{\#z}\, \hat{\#z} \ri
 \end{equation}
and
\begin{equation}
 \#\Gamma (\#r) =   \frac{f(r_s) v'_s \lec 1 + \les  1 - f(r_s) \ris \le v'_s \ri^2 \ric }{1 - \les  1 - f(r_s) \ris^2 \le v'_s \ri^2
 } \, \hat{\#x}. \l{vector_gamma}
\end{equation}

As we are interested in  a
   physically realizable metamaterial that represents the Alcubierre
   spacetime, the   limits
\begin{eqnarray}
&&\lim_{| \#r | \to 0} \; \=\gamma (\#r) = \hat{\#x}\, \hat{\#x} +
\les 1 - \le v'_s \ri^2 \ris  \le  \hat{\#y}\, \hat{\#y} +
\hat{\#z}\, \hat{\#z}
 \ri,  \qquad
\lim_{| \#r | \to  \infty} \; \=\gamma (\#r) = \=I
 ,\\
&& \lim_{| \#r | \to 0} \; \#\Gamma (\#r) = v'_s \, \hat{\#x},
\hspace{45mm} \lim_{| \#r | \to \infty}
 \#\Gamma (\#r) = \#0,
\end{eqnarray}
bear considerable promise.
Thus, the Tamm medium is like a gravitation-free  vacuum for large values of $| \#r |$
whereas its constitutive parameters remain  bounded at small values
of $| \#r |$.

 The  nontrivial constitutive parameters included in $ \=\gamma (\#r)$
and $ \#\Gamma (\#r)$, namely $\gamma_{22} \;(\equiv \gamma_{33})$
and $\Gamma_1$, are illustrated in Fig.~\ref{fig1} as functions of
$x$ and $y$ for $\sigma = 5$, $R = 1$ and $v'_s \in \lec 0.3, 0.6,
0.9 \ric$. The corresponding plots of $\gamma_{22} $ and
$\Gamma_{1}$
  versus
   $z$ are identical to those versus $y$. The constitutive parameters  $\gamma_{22}$
   and $\Gamma_1$ are continuous functions of $\#r$ for all values of $v'_s
   \in \les \, 0, 1 \ri$. Furthermore, the constitutive-parameter space is
   approximately partitioned into two disjoint regions with $\gamma_{22}$ and $\Gamma_1$ being approximately
   constant-valued in each. That is, we have
   \begin{itemize}
   \item[(i)] an inner
   region~---~which corresponds to the  warp bubble of Alcubierre
   spacetime~---~wherein $\gamma_{22} \approx 1  - \le v'_s \ri^2$ and
   $\Gamma_1 \approx  v'_s $, and
   \item[(ii)] an outer region wherein
   $\gamma_{22} \approx 1$ and $\Gamma_1 \approx 0$.
   \end{itemize}
   The inner
   region is shaped like a prolate spheroid whose major axis is aligned parallel to
   the $x$ axis. The prolate spheroid becomes increasingly elongated as the relative speed  $v'_s$ increases.

For the presentation of ray trajectories in Sec.~\ref{RT}, let us
introduce the semi-major axis length $a_M$ and semi-minor axis
length $a_m$ of the ellipse representing the inner region in the
$xy$ plane, defined via
\begin{equation} \l{axes}
\left.
\begin{array}{l}
 \displaystyle{\hat{\#x} \cdot \#\Gamma_{\,} ( a_M \hat{\#x}) =
\frac{1}{2}\, \hat{\#x} \cdot \#\Gamma_{} ( \#0 )} \vspace{8pt} \\
\displaystyle{\hat{\#x} \cdot \#\Gamma_{\,} ( a_m \hat{\#y})  =
\frac{1}{2} \,\hat{\#x} \cdot \#\Gamma_{} ( \#0 )}
\end{array}
\right\}.
\end{equation}

\section{Analysis of quasi-plane  waves}

As a precursor to our investigation of ray trajectories, we  first
consider  a quasi-plane wave whose electric and magnetic fields are of the form
\c{vanB}
\begin{equation}
\left.
\begin{array}{l}
 \#E (\#r,t) = {\rm Re} \lec \,\Eo(\#r) \exp \les i \le  \ko \#k\cdot \#r  - \omega
t \ri \ris\ric \vspace{6pt}
\\   \#H (\#r,t) = {\rm Re} \lec \, \Ho (\#r) \exp \les i \le  \ko \#k\cdot \#r - \omega t
\ri \ris\ric
\end{array}
\right\}.
  \l{qpw}
\end{equation}
 The quantities $\Eo (\#r)$ and $\Ho (\#r)$ in
eqs.~\r{qpw} are spatially varying, complex-valued amplitudes; $\omega$ is the angular
frequency; and  the   wavenumber $\ko = \omega \sqrt{\epso
\muo}$. Within the quasi--planewave regime  the relative wavevector
$\#k$ varies with $\#r$, but it is  convenient to omit the
dependency on $\#r$ in our notational representation of $\#k$.

The Maxwell curl postulates in combination with the constitutive
relations \r{CR2} and electromagnetic fields \r{qpw} yield the
nonhomogeneous vector differential equations
\begin{equation}
\left.
\begin{array}{l}
 \displaystyle{ \les \nabla \le \#k\cdot \#r \ri - \#\Gamma
(\#r) \ris \times \Eo (\#r) - \sqrt{\frac{\muo}{\epso}} \, \=\gamma
(\#r) \cdot \Ho (\#r)
= - \frac{1}{i \ko} \, \nabla \times \Eo (\#r) \vspace{12pt} }\\
\displaystyle{ \les \nabla \le \#k\cdot \#r \ri  - \#\Gamma (\#r)
\ris  \times \Ho (\#r) + \sqrt{\frac{\epso}{\muo}} \, \=\gamma (\#r)
\cdot \Eo (\#r) = - \frac{1}{i \ko} \, \nabla \times \Ho (\#r)}
\end{array}
\right\} \l{m2}.
\end{equation}
Under the geometric-optics approximation, the constitutive
parameters are assumed to vary only very slowly over the distance
of a wavelength. Thus,  $ \nabla \le \#k\cdot \#r \ri \approx \#k$
and the right sides of eqs.~\r{m2} are approximately null-valued.
Hence eqs.~\r{m2} reduce to \c{MLS_NJP}
\begin{equation}
\lec \les \det \=\gamma (x, y) - \#p   \cdot \=\gamma (x, y ) \cdot
 \#p   \,\ris \=I + \#p  \, \#p  \cdot \=\gamma (x, y) \ric \cdot \Eo (\#r) =
 \#0 \,, \l{e1}
\end{equation}
wherein the vector $\#p  =  \#k  - \#\Gamma (x, y)$ is introduced.
The existence of   nonzero solutions to eq.~\r{e1} imposes the
condition
\begin{equation}
\mathcal{H}  \equiv  \det \=\gamma (x, y) - \#p   \cdot \=\gamma (x,
y ) \cdot
 \#p  = 0. \l{disp}
\end{equation}

Equation \r{disp} represents the dispersion relation from which
 the  magnitude $k$ of the relative wavevector $\#k$ may be extracted as follows.
Writing $\#k = k$ $ ( \sin
\theta \,\cos \phi\,\hat{\#x} + \sin \theta \,\sin \phi\, \hat{\#y}+ \cos \theta\,\hat{\#z}
)$, we see that the left side of eq. \r{disp} is quadratic in $k$; hence,
\begin{equation}
k \equiv k^\pm (\theta, \phi) = \frac{-b \pm \sqrt{b^2 - 4 a c }}{2
a}, \l{kroot}
\end{equation}
where the coefficients
\begin{eqnarray}
a &=&  \frac{\les 1 - \le v'_s \ri^2 \ris\le \cos^2 \theta + \sin^2
\theta \sin^2 \phi \ri + \lec 1 - \les 1 - f(r_s) \ris^2 \le v'_s
\ri^2 \ric \sin^2 \theta \cos^2 \phi}{1 - \les 1 - f(r_s) \ris^2 \le
v'_s \ri^2
 } ,\\
b &=& - \frac{2 \ko v'_s f(r_s) \lec 1 + \les 1 - f(r_s) \ris \le
v'_s \ri^2 \ric \sin \theta \cos \phi}{1 - \les 1 - f(r_s) \ris^2
\le v'_s \ri^2
 } , \\
c &= & - \frac{ \ko^2 \le 1 -  \lec 1 +  f(r_s) \les 2- f(r_s)  \le
v'_s \ri^2 \ris \ric \le v'_s \ri^2 \ri}{1 - \les 1 - f(r_s) \ris^2
\le v'_s \ri^2
 }.
\end{eqnarray}
Let us note that $k^+ (\pi - \theta, \pi + \phi) = - k^- (\theta,
\phi)$; i.e., $k^-$ is the wavenumber for a quasi-plane wave travelling in
the opposite direction to the quasi-plane wave with wavenumber $k^+$. This
observation is a manifestation of the unirefringence of  vacuum
\c{LMcpl}.

The discriminant term $b^2 - 4 ac$ in eq.~\r{kroot} can have a
negative value. Therefore the relative wavenumber
 $k$ may be complex-valued with nonzero imaginary part, in spite of the fact that
all the coefficients of the dispersion relation \r{disp} have
real values.
 However,
 $\mbox{Im} \lec  k \ric \neq 0$ is indicative of evanescent waves which we
  exclude from our study of ray trajectories.

  The partition of the $k$--phase space into
  a propagating-wave regime and an evanescent-wave regime
   is illustrated in Fig.~\ref{fig2}, wherein the
directions of $\#k$  for which $\mbox{Im} \lec  k \ric \neq 0$ are
represented
 at  the coordinate origin for $v'_s \in \lec 0.60, 0.62, 0.70,  0.90 \ric$.
We choose the parameter values  $\sigma  =5$ and $R= 1$ for these
plots.
 Only the directions in one octant of the unit sphere
need be displayed, because of symmetry. At relative speeds
$v'_s \leq 0.6 $, the  relative wavenumbers are  wholly real-valued for
all propagation directions. As $v'_s$ increases just beyond $0.6$,
relative wavenumbers with nonzero imaginary parts emerge  for $\#k$
directed in the $yz$ plane. As $v'_s$ increases further, $\mbox{Im}
\lec  k \ric \neq 0$ occurs  at increasingly larger values of
$\hat{\#x} \cdot \#k$; in the limit $v'_s \to 1$, we find that
$\mbox{Im} \lec  k \ric \neq 0$ occurs for all directions of propagation. The same trend is observed at locations throughout the
inner region of the constitutive-parameter space referred to in our
discussion of Fig.~\ref{fig1}. In the outer region, however,  $k$ is everywhere
real-valued for all propagation directions.

\section{Ray trajectories} \l{RT}

 A convenient Hamiltonian function for our ray-tracing study is
 provided by
the scalar quantity $\mathcal{H} $ introduced in eq.~\r{disp}. We
parameterize
 the ray trajectories  in terms of $\tau$ via  $\#r
(\tau)$; similarly, the parameterization  $\#k (\tau)$ is used for
the relative wavevector. The ray trajectories are thus
governed by the coupled vector differential equations \c{Kline,
Sluij2}
\begin{equation}
\left. \begin{array}{l} \displaystyle{
 \frac{d \#r}{d \tau} =  \nabla_{\#k}
 \mathcal{H}} \vspace{10pt} \\
\displaystyle{\frac{d \#k}{d \tau} = -  \nabla_{\#r}
 \mathcal{H}}
\end{array} \right\}, \l{odes}
\end{equation}
 wherein  the
shorthand $\nabla_{\#q} \equiv\hat{\#x}\,
\partial/ \partial q_x + \hat{\#y} \,\partial/ \partial q_y+\hat{\#z}\,\partial/ \partial
q_z $ for $\#q =  q_x\,\hat{\#x}+q_y\,\hat{\#y}+q_z \,\hat{\#z}$ is employed.
 Once appropriate initial conditions $\#r(0)$ and $\#k(0)$ have been specified, the system
\r{odes} can be solved for $\#r( \tau)$ and $\#k (\tau)$ using
standard numerical methods---e.g., the Runge--Kutta method
\c{Sluij2}.

That  the direction of  a ray trajectory, as provided by the
direction of $ \nabla_{\#k} \mathcal{H} $, is aligned  with the
direction of energy flux (quantified as the
 time-averaged Poynting vector)
was demonstrated previously for a general Tamm medium, under the
proviso that $\=\gamma (\#r)$ is either positive-definite or
negative-definite  \c{MLS_NJP,THA}. It is clear from
eq.~\r{gamma_dyadic} that both distinct eigenvalues of $\=\gamma (\#r)$,
namely $1$ and $ \frac{1 - \le v'_s \ri^2}{1 - \les  1 - f(r_s)
\ris^2 \le v'_s \ri^2
 } $, are positive-valued. Hence, the direction of energy flux is indeed
 aligned with the ray trajectories deduced from
eqs.~\r{odes}.

\begin{figure}[!h]
\centering \psfull \epsfig{file=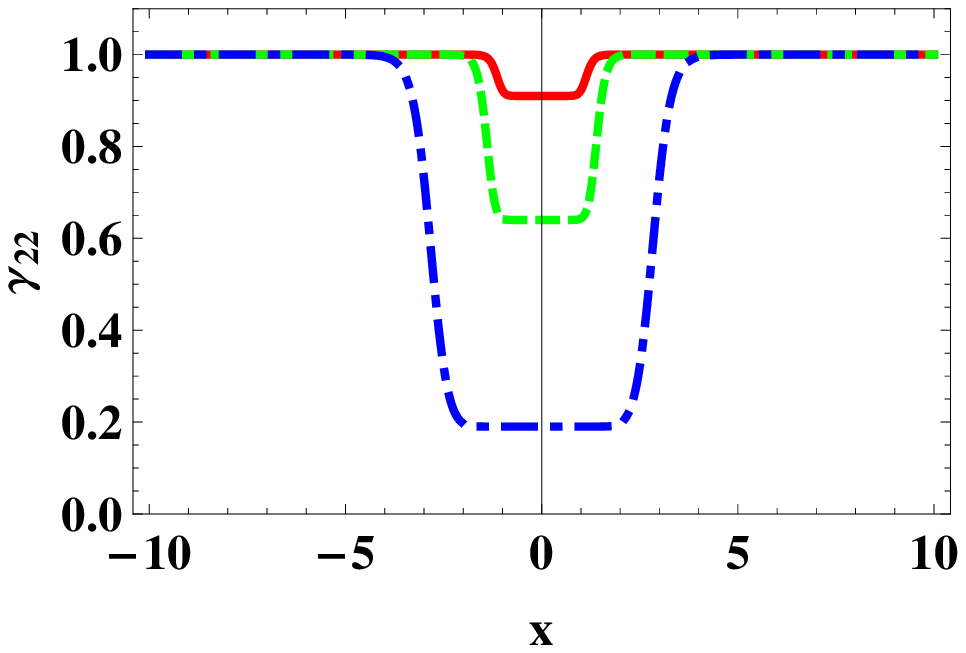,width=3.2in}
\epsfig{file=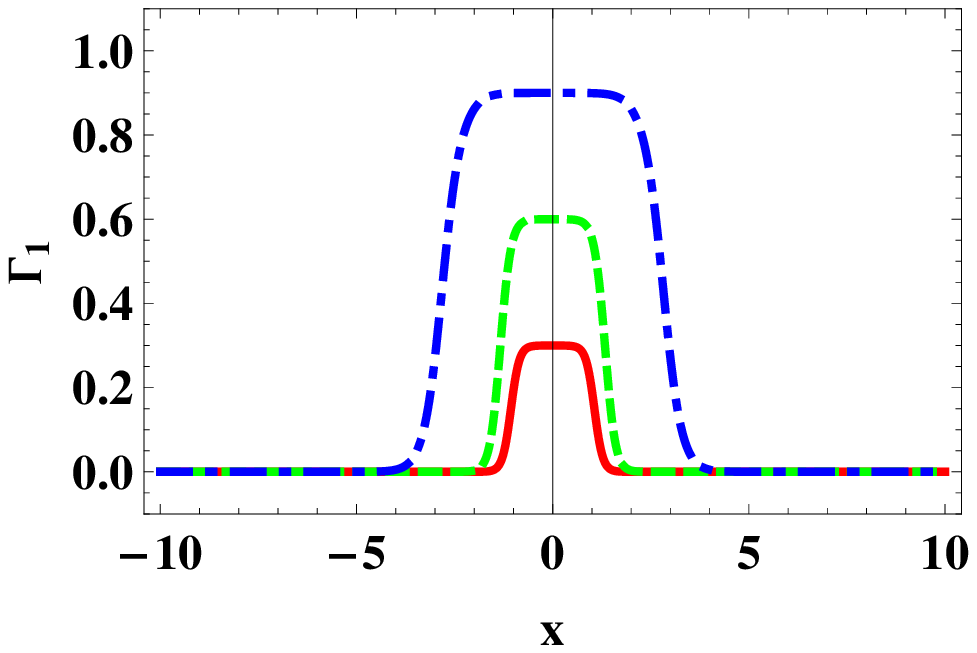,width=3.2in} \\
\epsfig{file=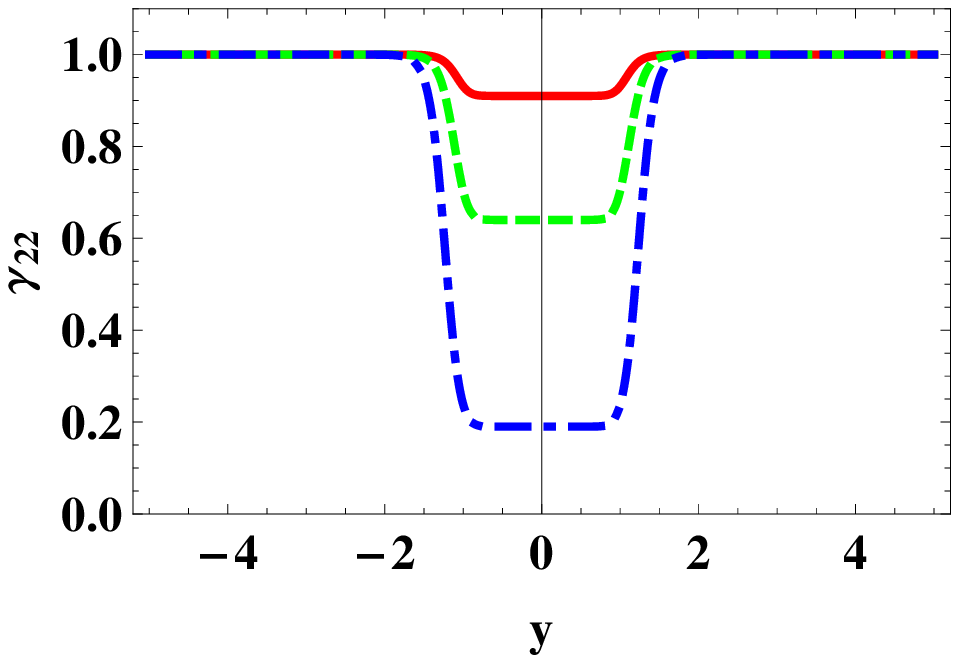,width=3.2in}
\epsfig{file=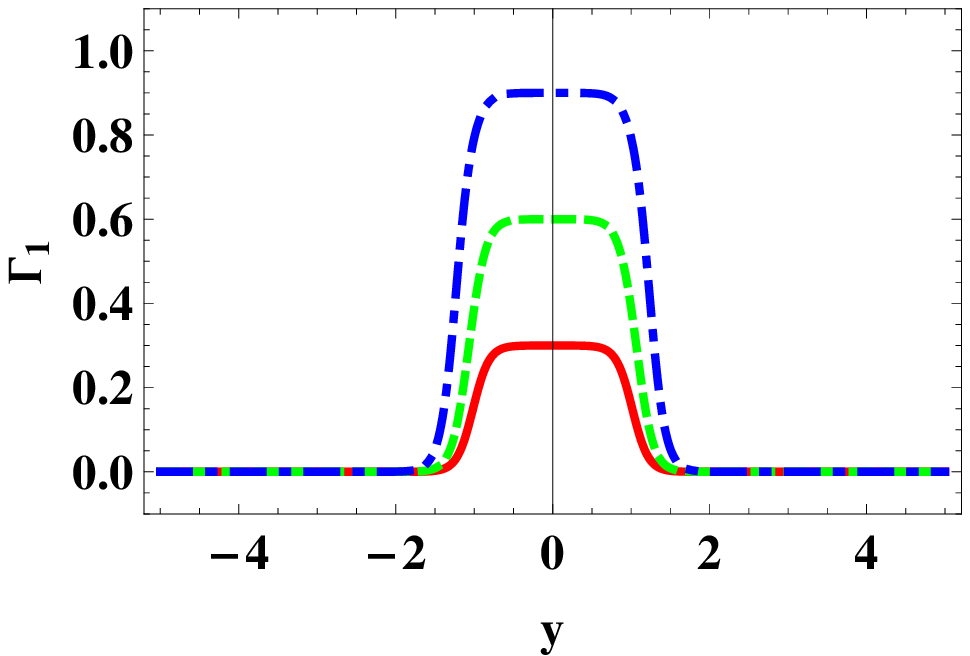,width=3.2in}
 \caption{The constitutive parameters $\gamma_{22} $ and $\Gamma_{1}$
  plotted versus $x$  for $y = z  =0$, and versus  $y$ for $x = z  =0$. Parameter values:  $\sigma  =5$, $R= 1$ and
  $v'_s = 0.9$ (blue, broken dashed curves), $0.6$ (green, dashed curves) and $0.3$ (red, solid curves).
} \label{fig1}
\end{figure}

\begin{figure}[!h]
\centering \psfull
 \epsfig{file=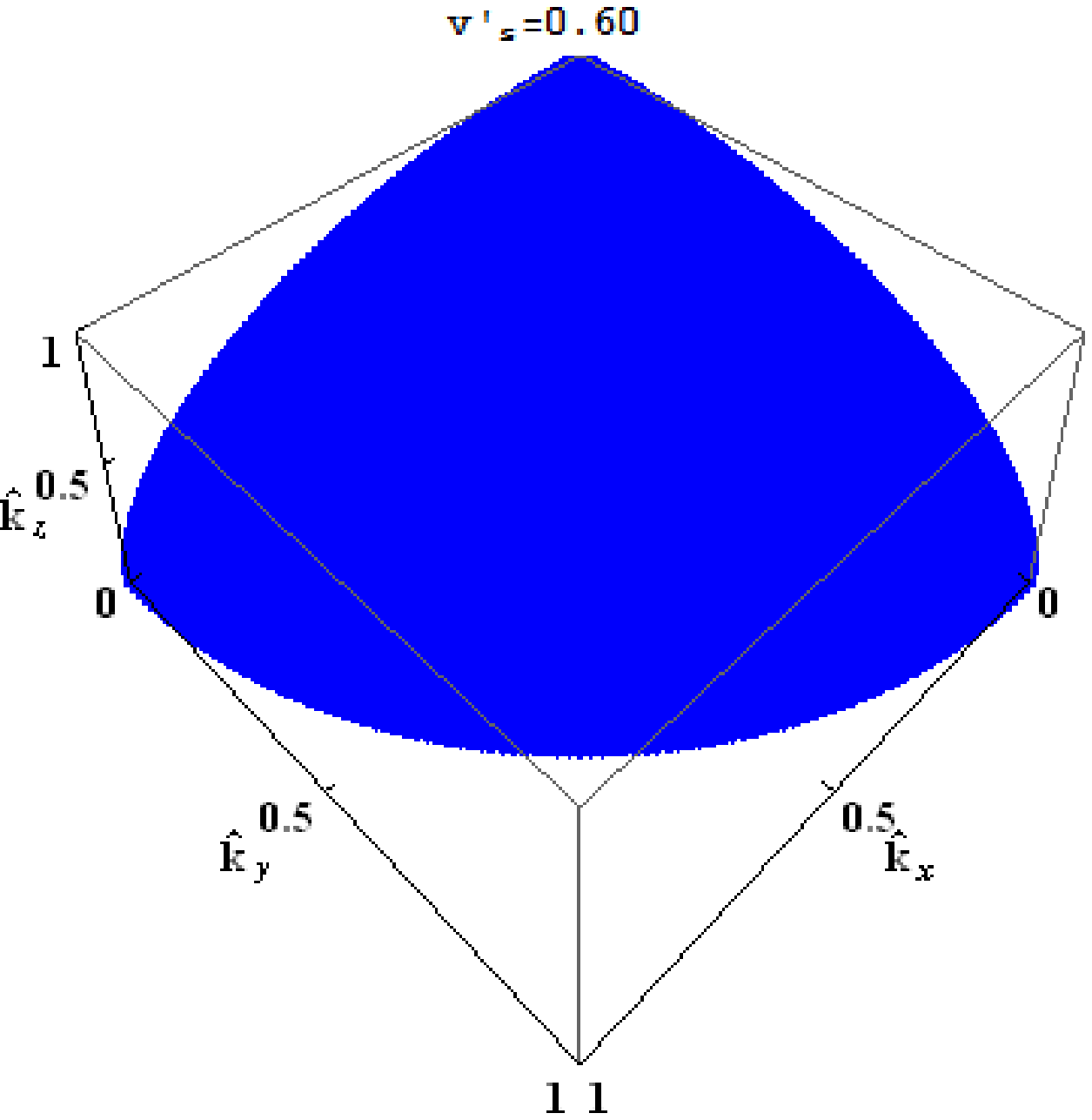,width=3.2in}
\epsfig{file=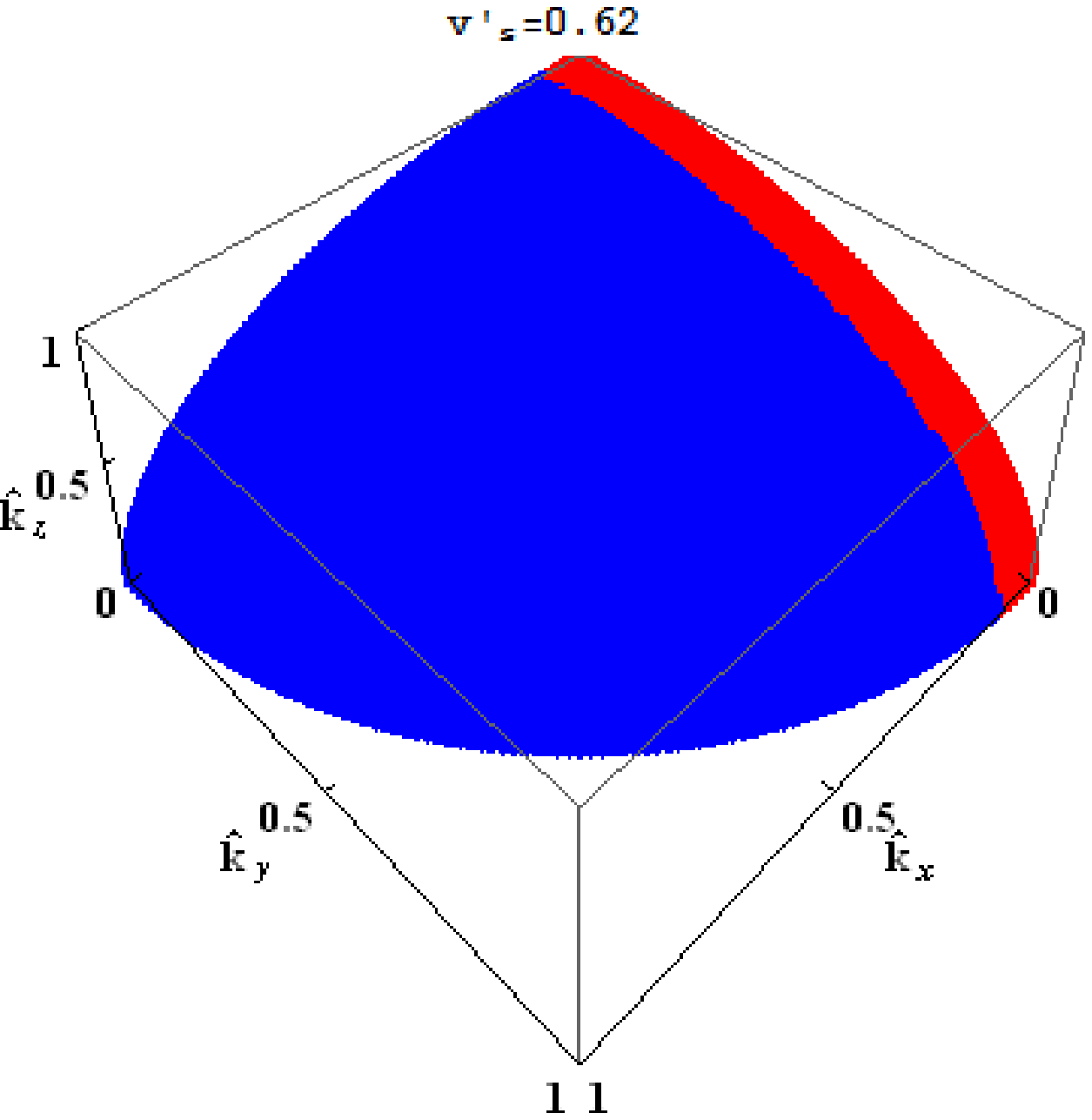,width=3.2in} \\
\epsfig{file=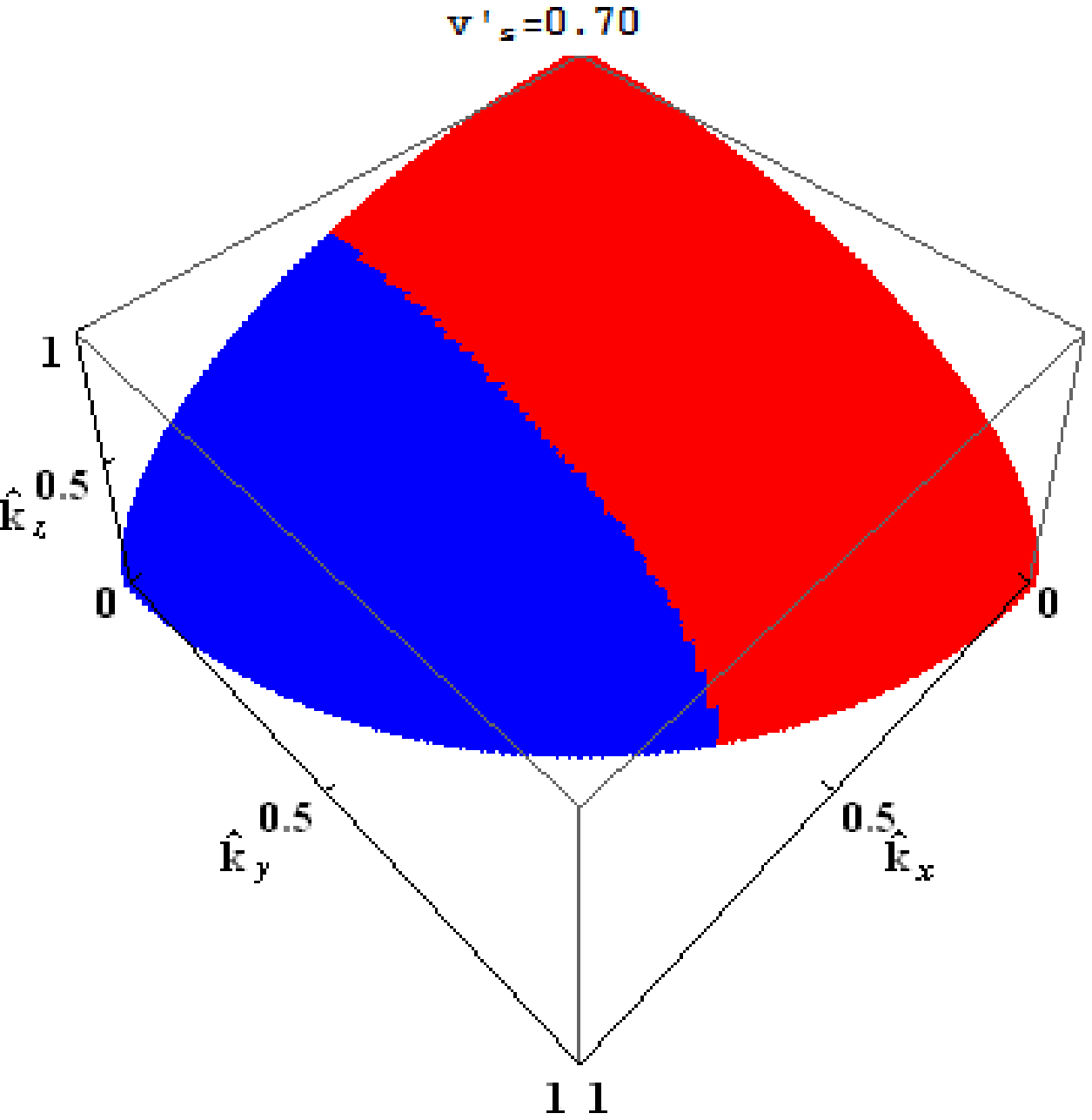,width=3.2in}
\epsfig{file=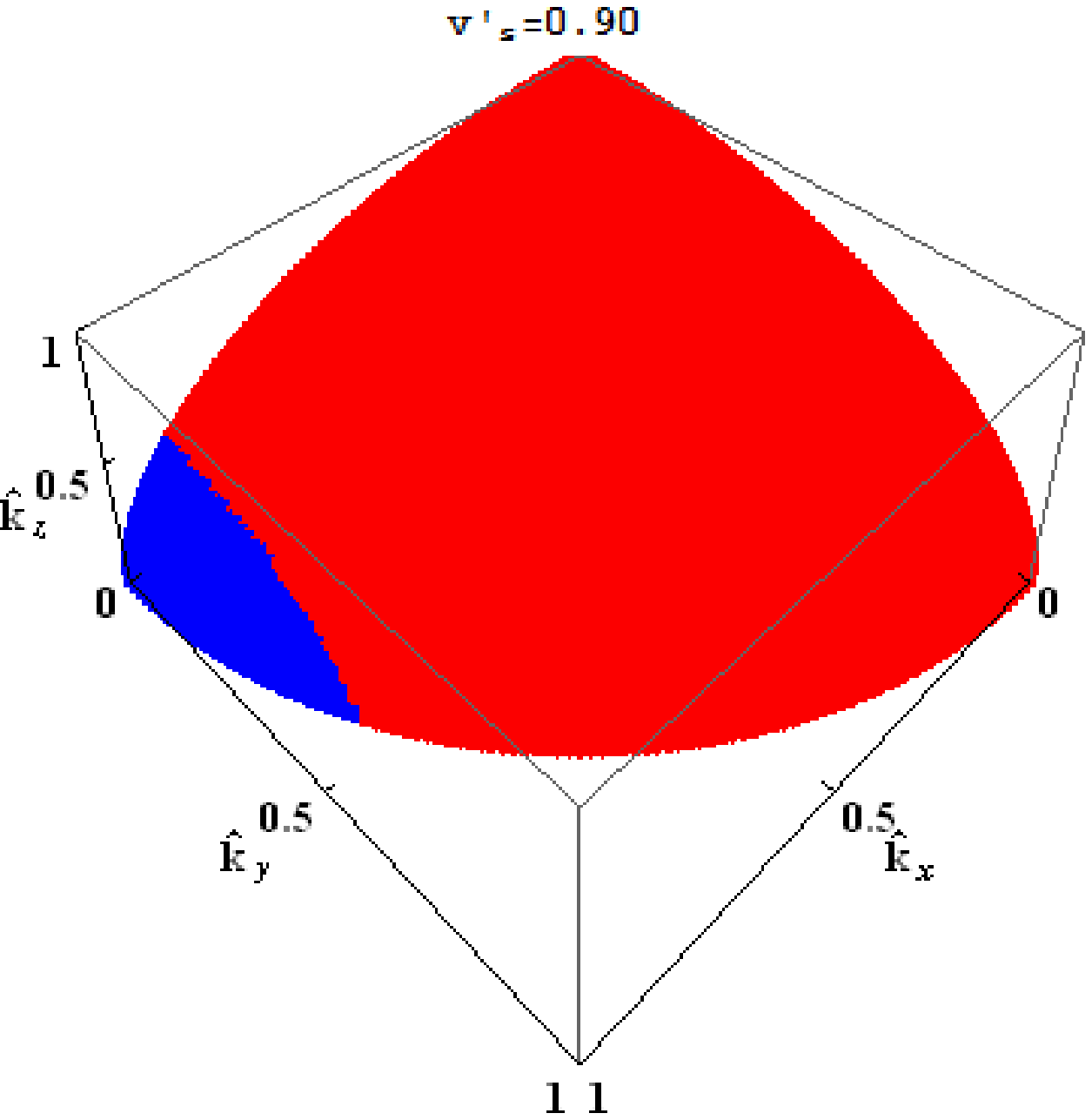,width=3.2in}
 \caption{Maps illustrating the
directions of $\hat{\#k} = \#k/k =
\hat{k}_x\,\hat{\#x}+\hat{k}_y\,\hat{\#y}+\hat{k}_z\,\hat{\#z} $
 for
which $\mbox{Im} \lec k \ric = 0$ (blue) and $\mbox{Im} \lec k \ric
\neq 0$ (red),
 at the coordinate origin for $v'_s \in \lec 0.60, 0.62, 0.70, 0.90 \ric$. Parameter values:  $\sigma  =5$ and $R= 1$. } \label{fig2}
\end{figure}

\begin{figure}[!h]
\centering \psfull \epsfig{file=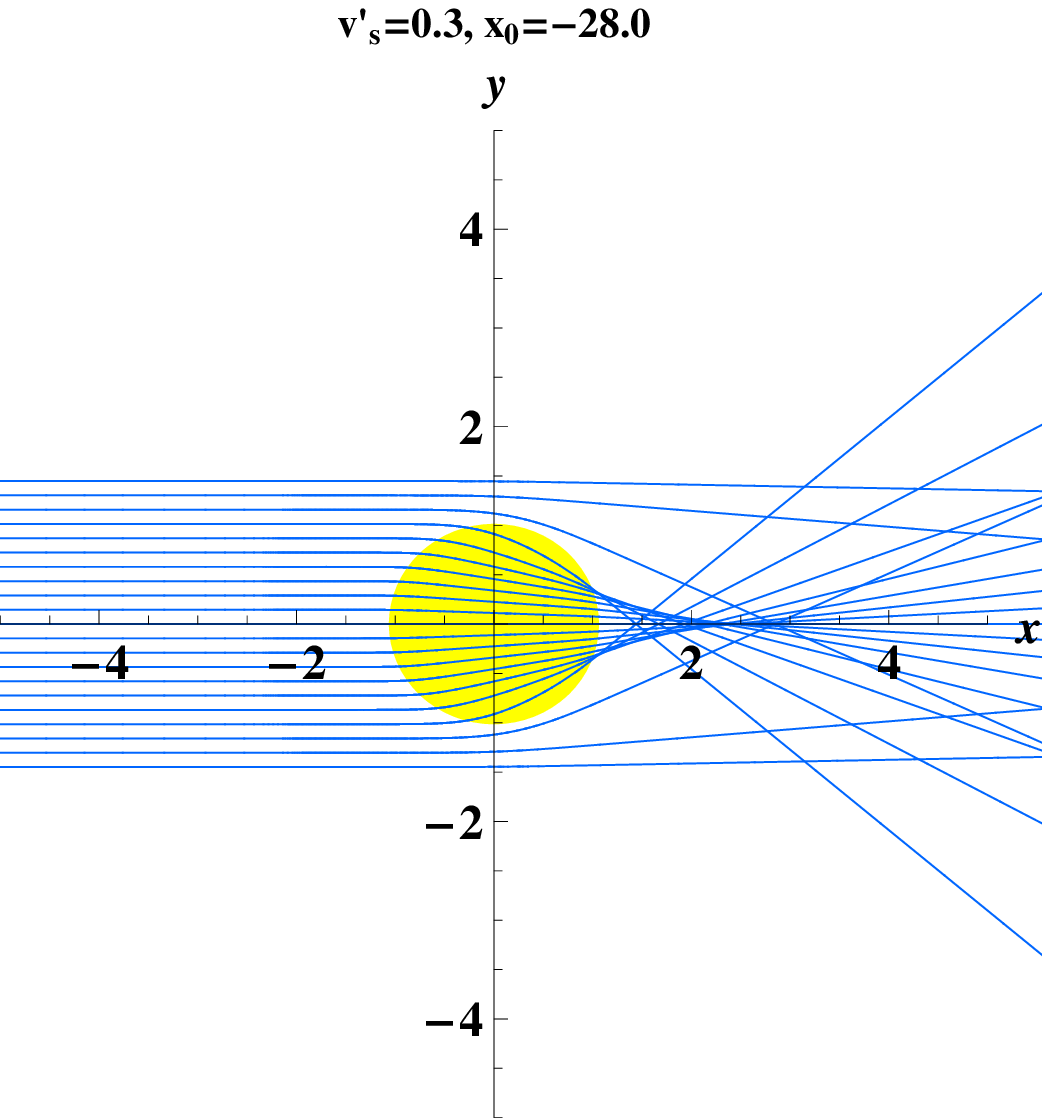,width=3.2in}
\epsfig{file=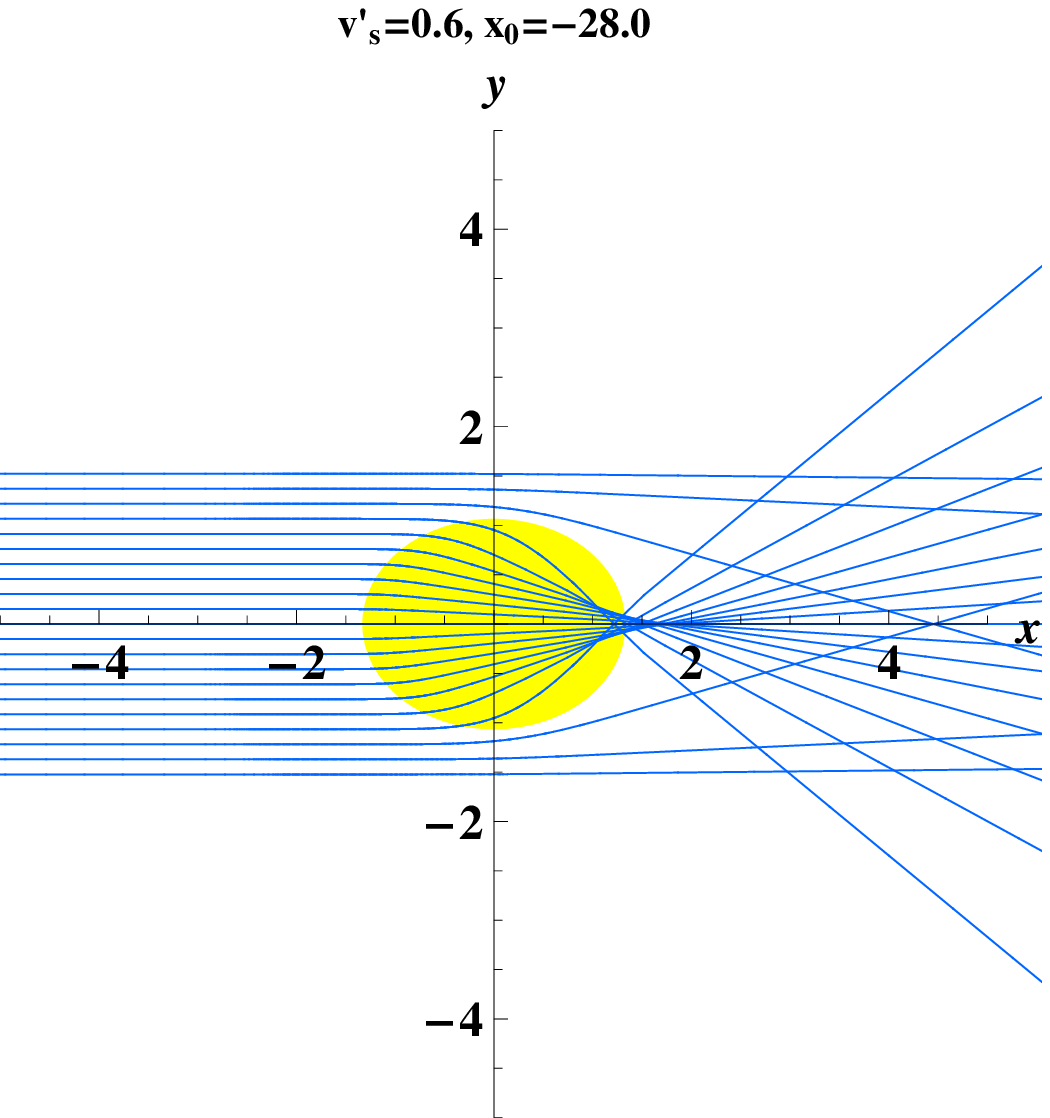,width=3.2in} \\
\epsfig{file=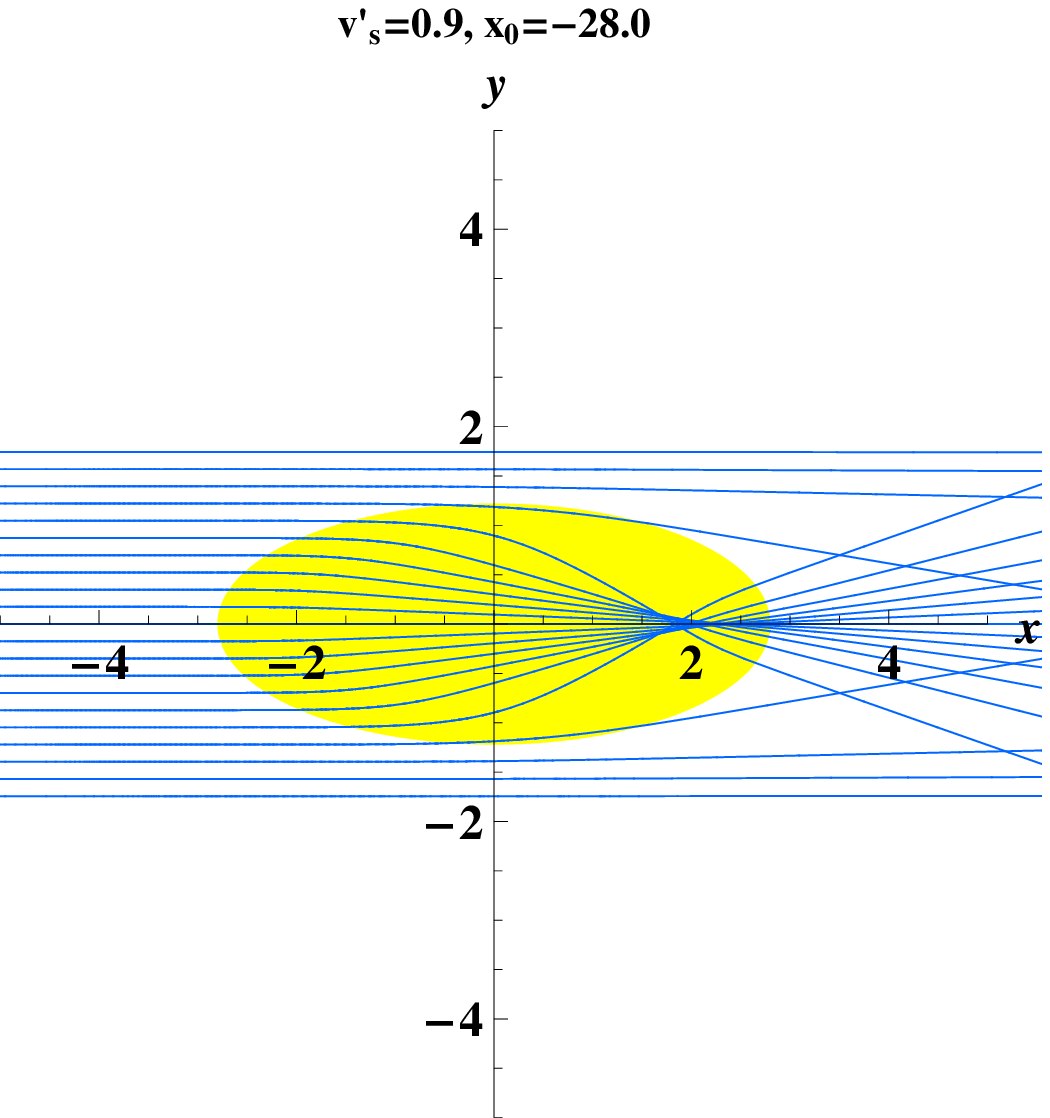,width=3.2in}
 \caption{Trajectories for rays in the $xy$ plane,
 initially parallel to the $x$ axis and equally spaced. That is,
  $\#r(0) =x_0\,\hat{\#x}+y_0\,\hat{\#y}$ with fixed $x_0 = -28 $  and
  $-1.5\, a_m
< y_0 < 1.5 \, a_m$,  $\#k (0) = \hat{\#x}$ and  $v'_s \in \lec 0.3,
0.6, 0.9 \ric$. Parameter values: $\sigma = 5$ and $R = 1$. }
\label{fig3}
\end{figure}

We begin our presentation of ray trajectories by considering an
array of rays in the $xy$ plane, initially parallel to the $x$ axis
and equally spaced. That is, we take $\#r(0) =  x_0\, \hat{\#x}+
y_0\,\hat{\#y}$ with $x_0$ fixed and $-1.5 \,a_m < y_0 < 1.5 \,a_m$.
As in Figs.~\ref{fig1} and \ref{fig2}, we set $\sigma = 5$ and $R =
1$. In Fig.~\ref{fig3}, ray trajectories are shown for
  $x_0 < 0$, $\#k (0) = \hat{\#x}$, and  $v'_s
\in \lec 0.3, 0.6, 0.9 \ric$. The inner
 region of the constitutive-parameter space referred to in our
discussion of Fig.~\ref{fig1} is shown as a shaded (yellow) ellipse
(with semi-major axis length $a_M$ and semi-minor axis length $a_m$,
per eqs.~\r{axes}), which becomes more eccentric at larger values of
$v'_s$. The inner region is seen to have a focusing effect, with the
focus lying on the $+x$ axis. Furthermore, the focus shifts towards
the coordinate origin as the relative speed $v'_s$ increases.

\begin{figure}[!h]
\centering \psfull \epsfig{file=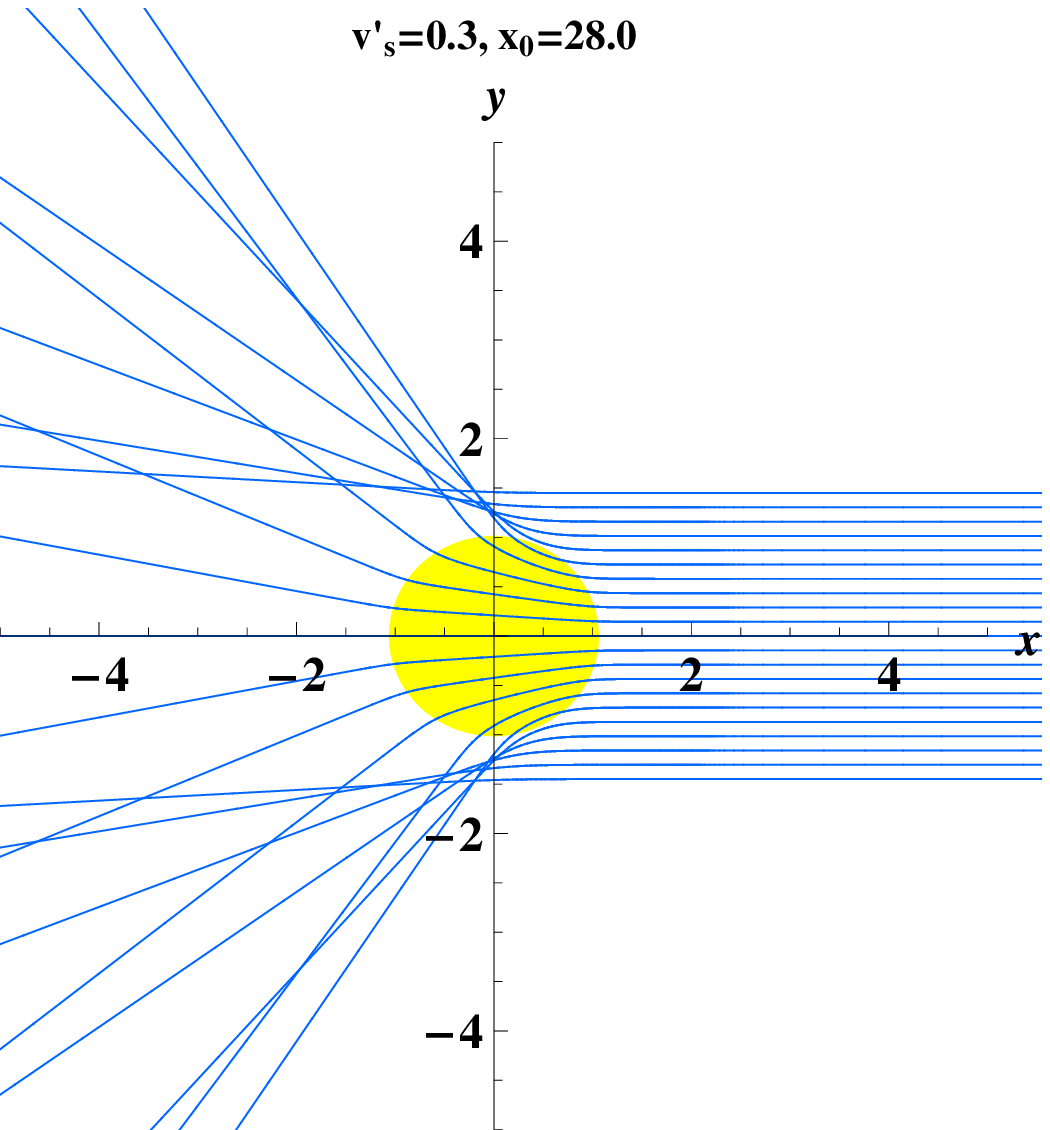,width=3.2in}
\epsfig{file=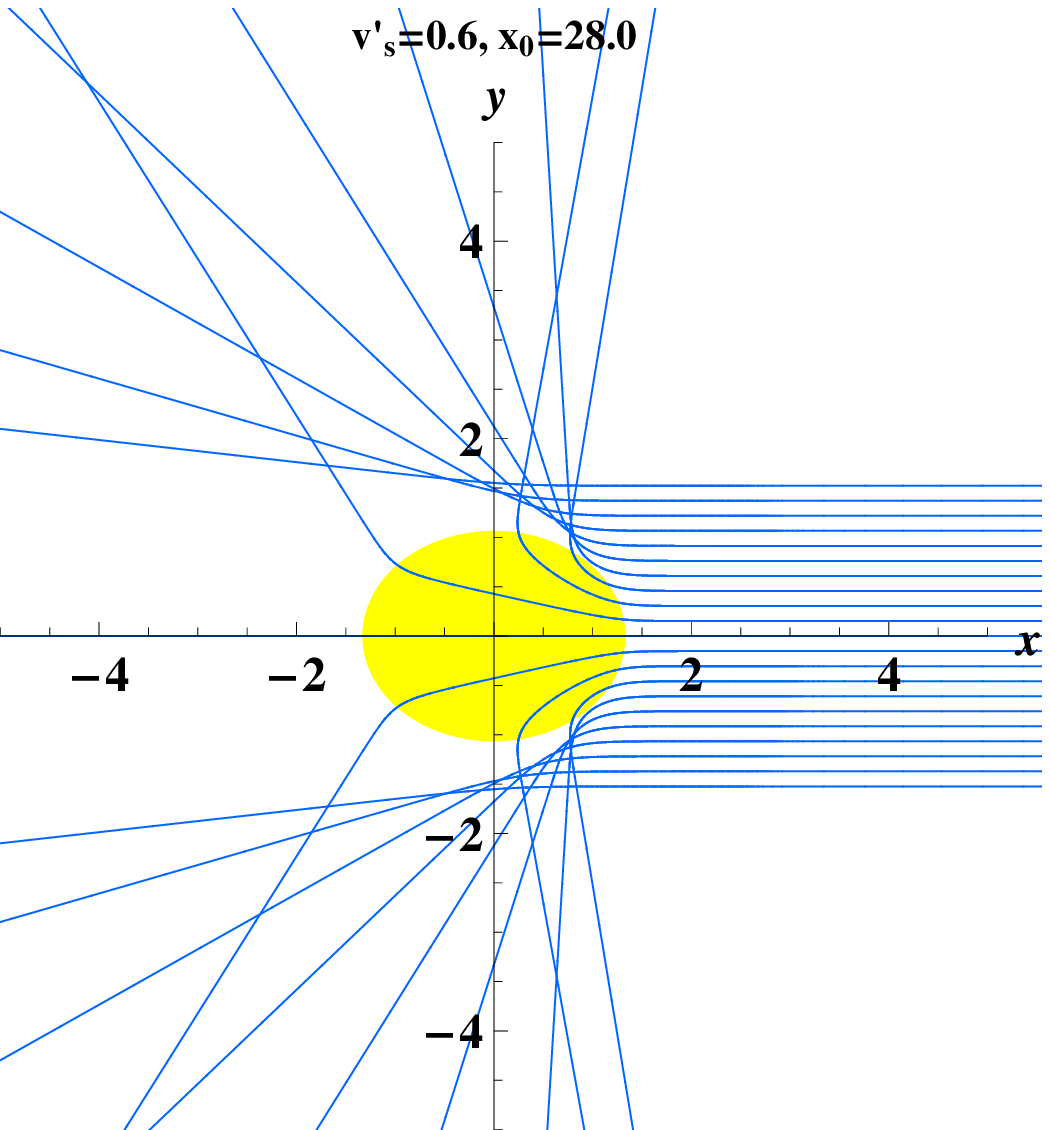,width=3.2in} \\
\epsfig{file=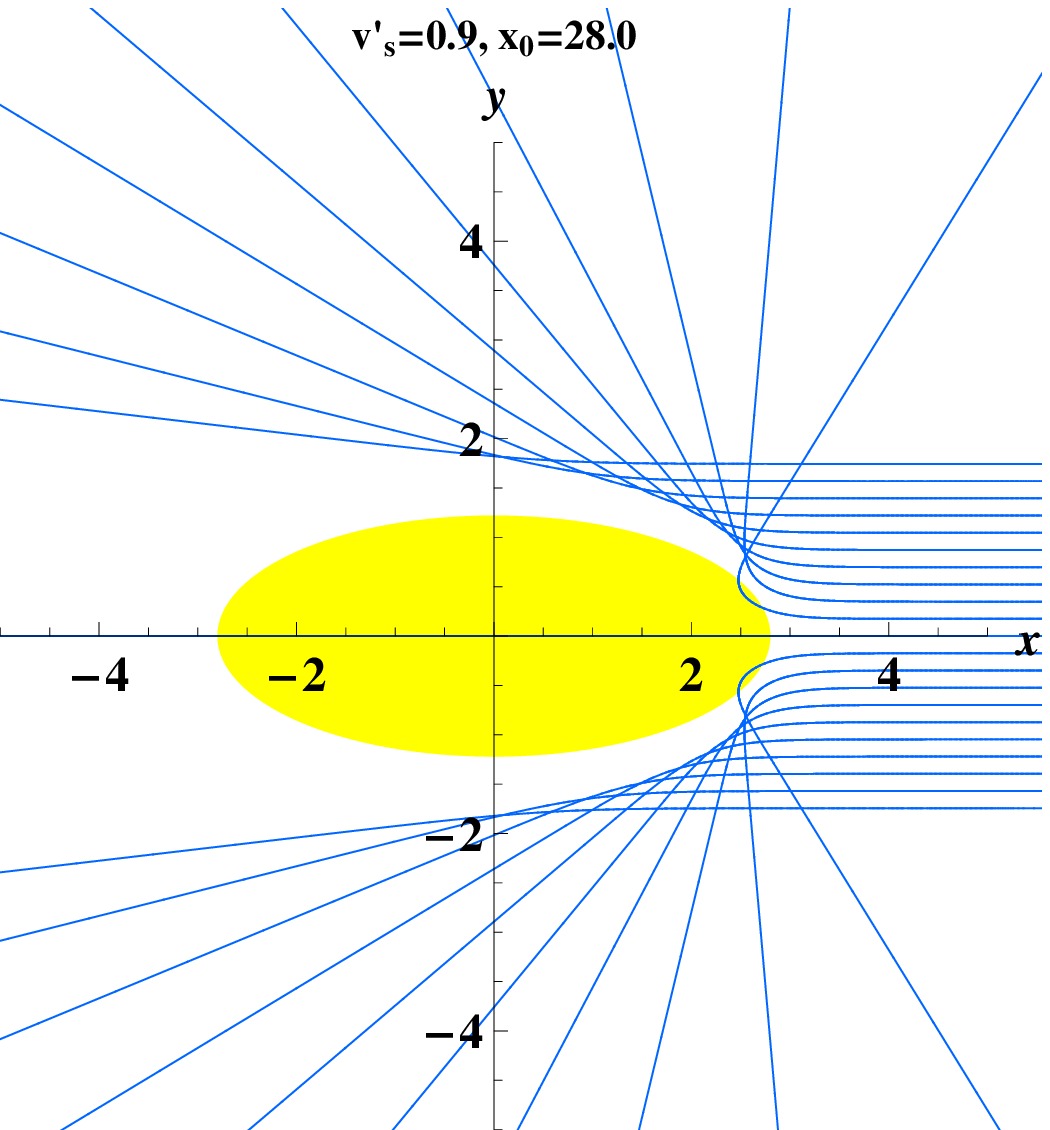,width=3.2in}
 \caption{As Fig.~\ref{fig3} except that
   $x_0  = 28 $  and   $\#k (0) = -\hat{\#x}$. } \label{fig4}
\end{figure}

That the Tamm medium is not reciprocal in the Lorentz sense \c{EAB}
is vividly illustrated by a comparison of Figs.~\ref{fig3} and
\ref{fig4}. The scenario represented in Fig.~\ref{fig4} is the same
as that of Fig.~\ref{fig3} except that $x_0 > 0$ and $\#k (0) =
-\hat{\#x}$. Quite unlike Fig.~\ref{fig3}, there is no evidence of
focusing by the inner region in Fig.~\ref{fig4}. On the contrary,
rays appear to diverge as they pass through the inner region at low
values of $v'_s$, while rays are almost entirely excluded from the
inner region altogether at $v'_s = 0.9$.

\begin{figure}[!h]
\centering \psfull \epsfig{file=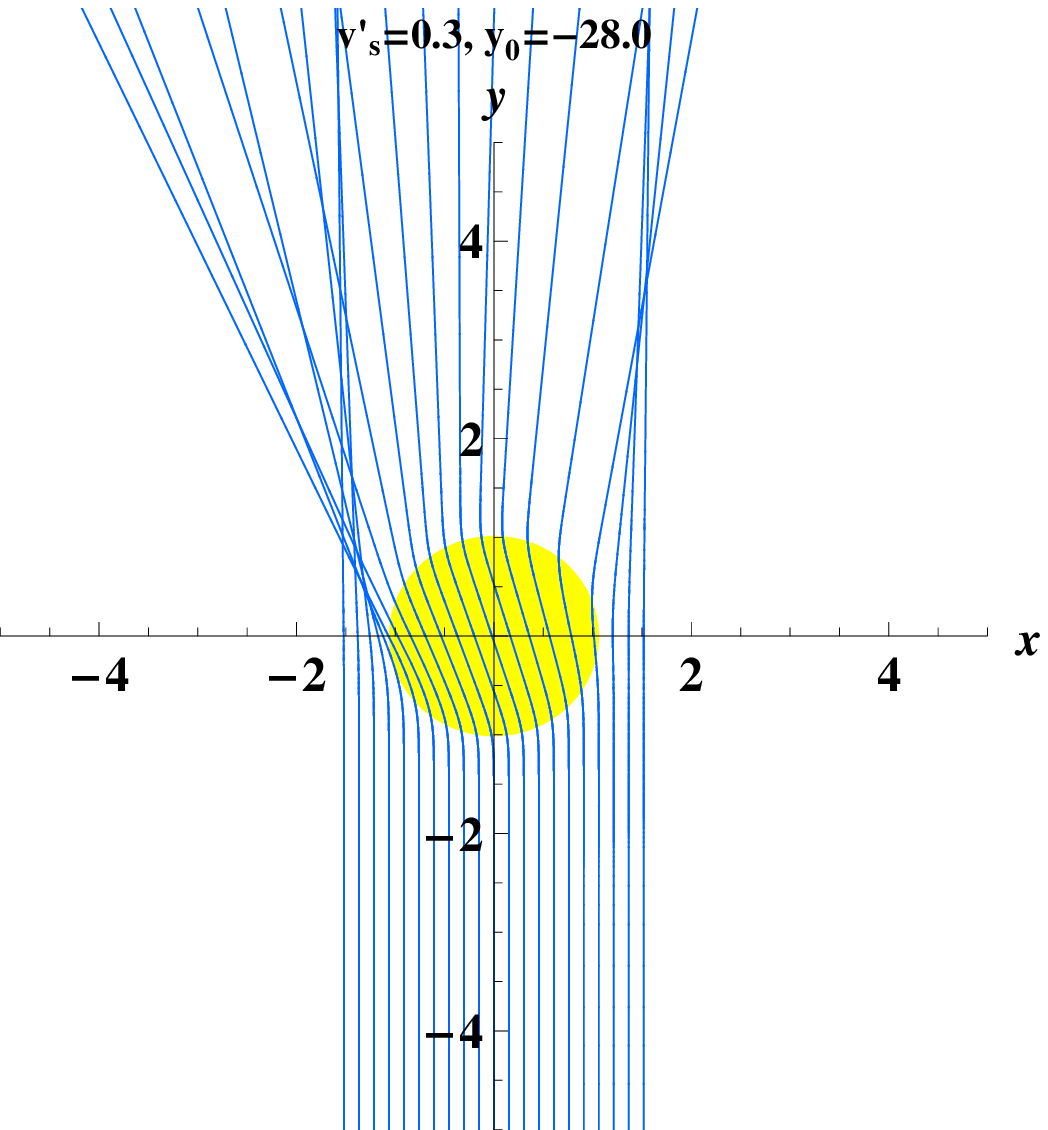,width=3.2in}
\epsfig{file=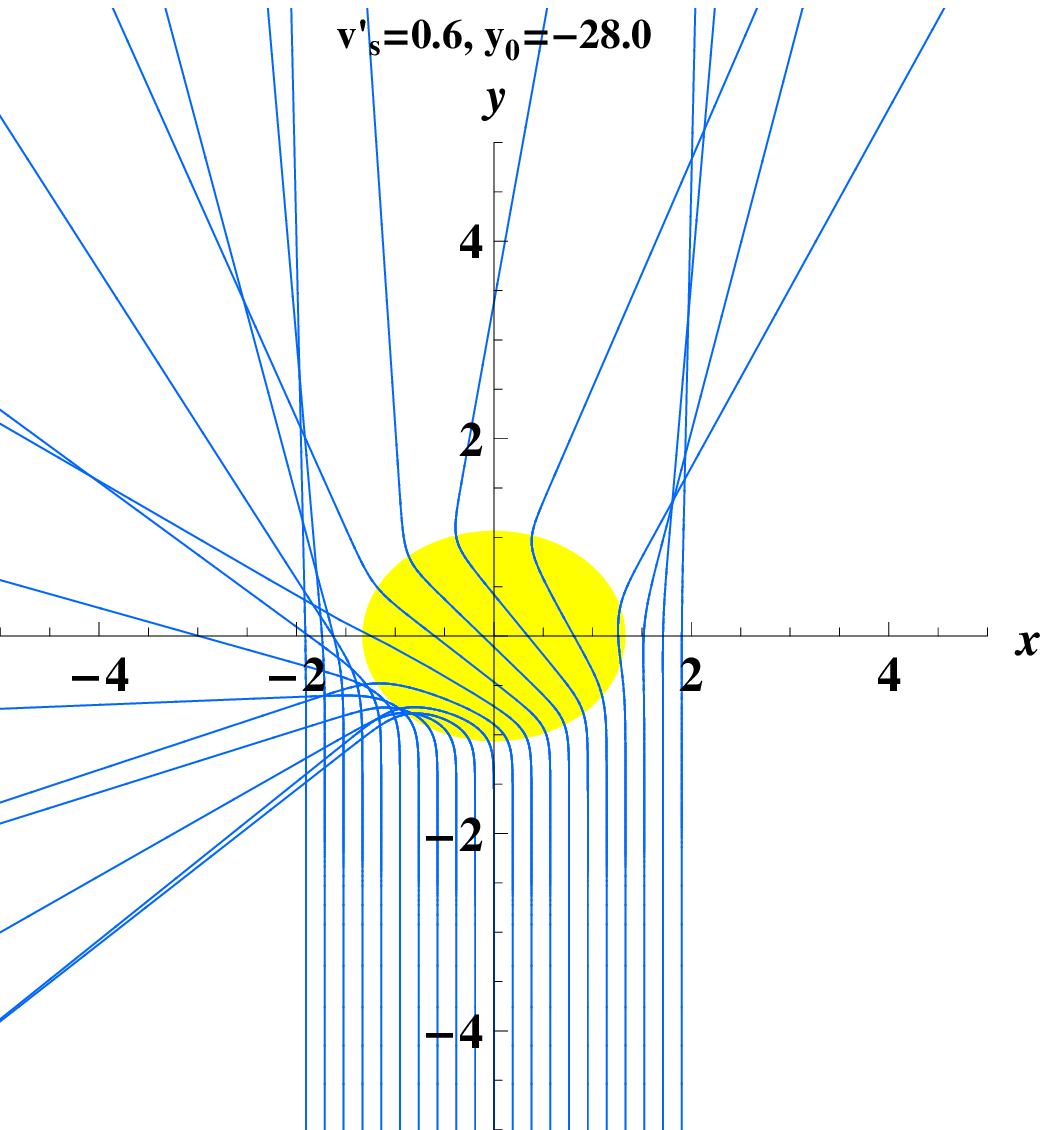,width=3.2in} \\
\epsfig{file=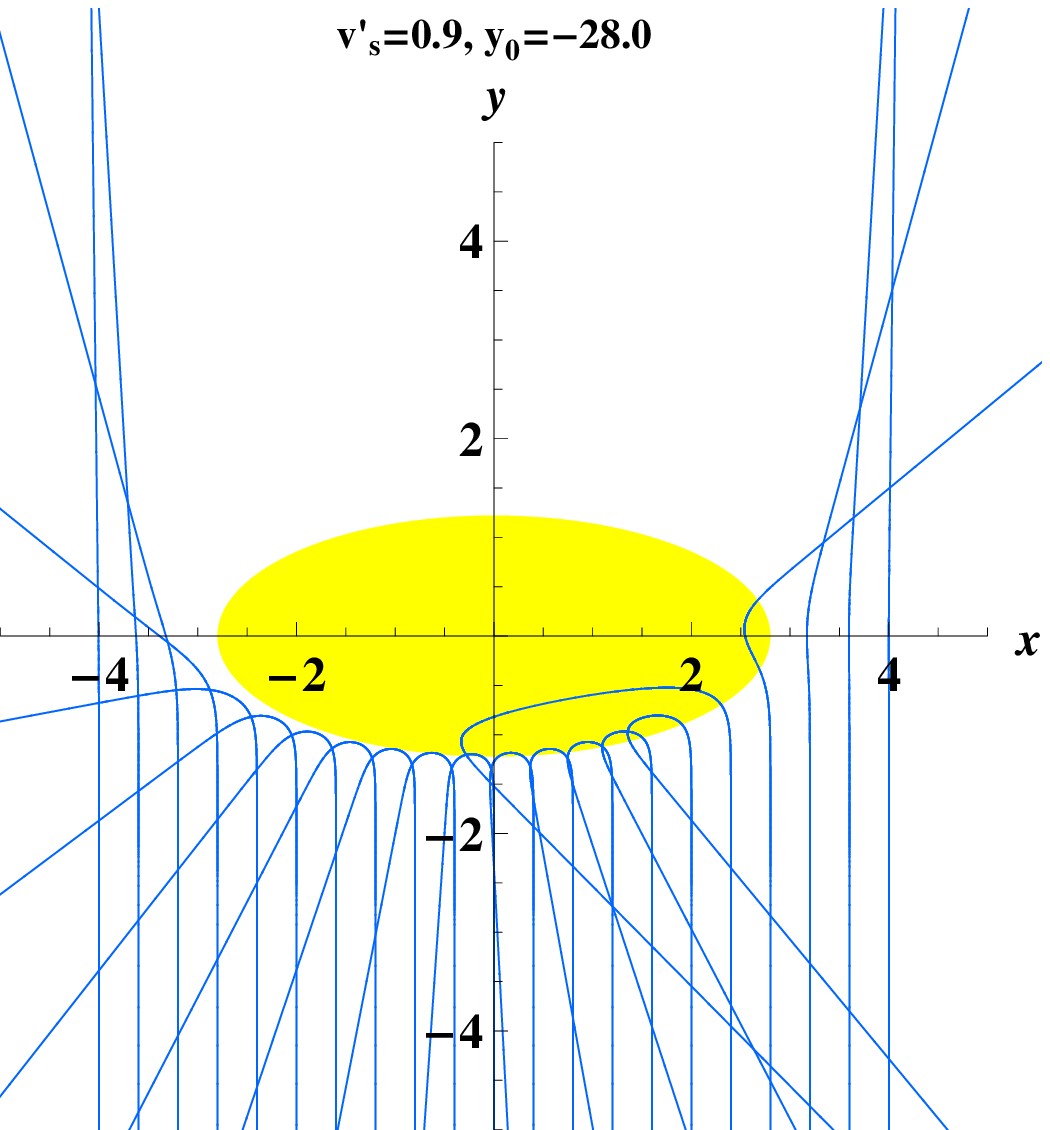,width=3.2in}
 \caption{As Fig.~\ref{fig3} except that
   $\#r(0) = x_ 0\,\hat{\#x}+ y_0\,\hat{\#y}$ with  $-1.5 \, a_M < x_0 < 1.5 \, a_M$  and fixed  $ y_0 = -28$,
   and
     $\#k (0) = \hat{\#y}$ }. \label{fig5}
\end{figure}

Further insight into  the absence of Lorentz-reciprocity of the Tamm
medium  is provided in Fig.~\ref{fig5} wherein ray trajectories
initially parallel to the $y$ axis and equally spaced are presented.
The initial relative wavevector for these trajectories is $\#k (0) =
\hat{\#y}$, and  we have set $\#r(0) =x_0\, \hat{\#x}+
y_0\,\hat{\#y}$ with fixed $y_0 < 0$  while $-1.5\, a_M < x_0 <
1.5\, a_M$. As for Figs.~\ref{fig1}--\ref{fig4},  $\sigma = 5$ and
$R = 1$. The plots in Fig.~\ref{fig5} are clearly asymmetric with
respect to the $y$ axis, and the ray trajectories become
progressively excluded from the inner region as $v'_s$ increases.

\begin{figure}[!h]
\centering \psfull \epsfig{file=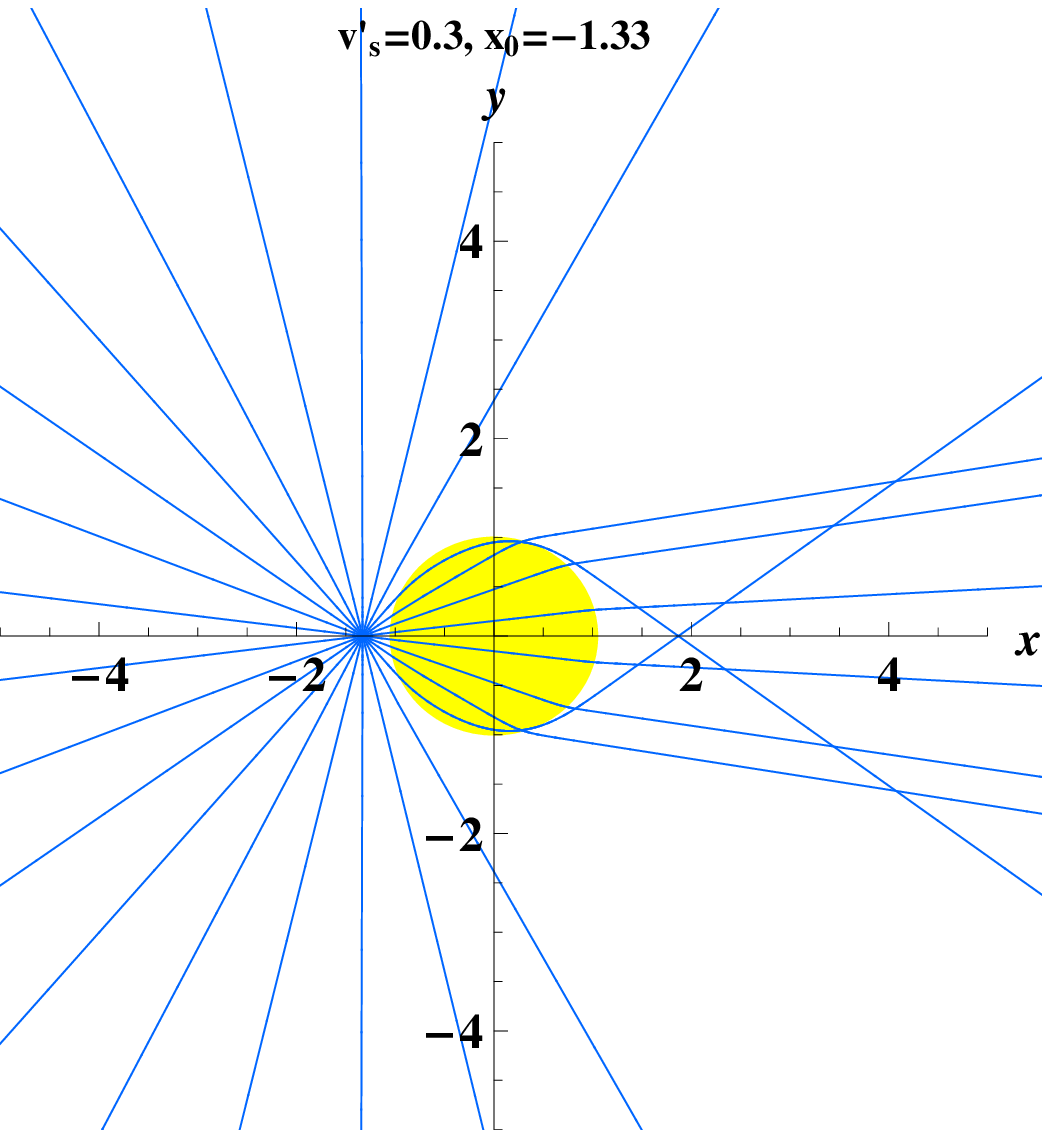,width=1.9in}
\epsfig{file=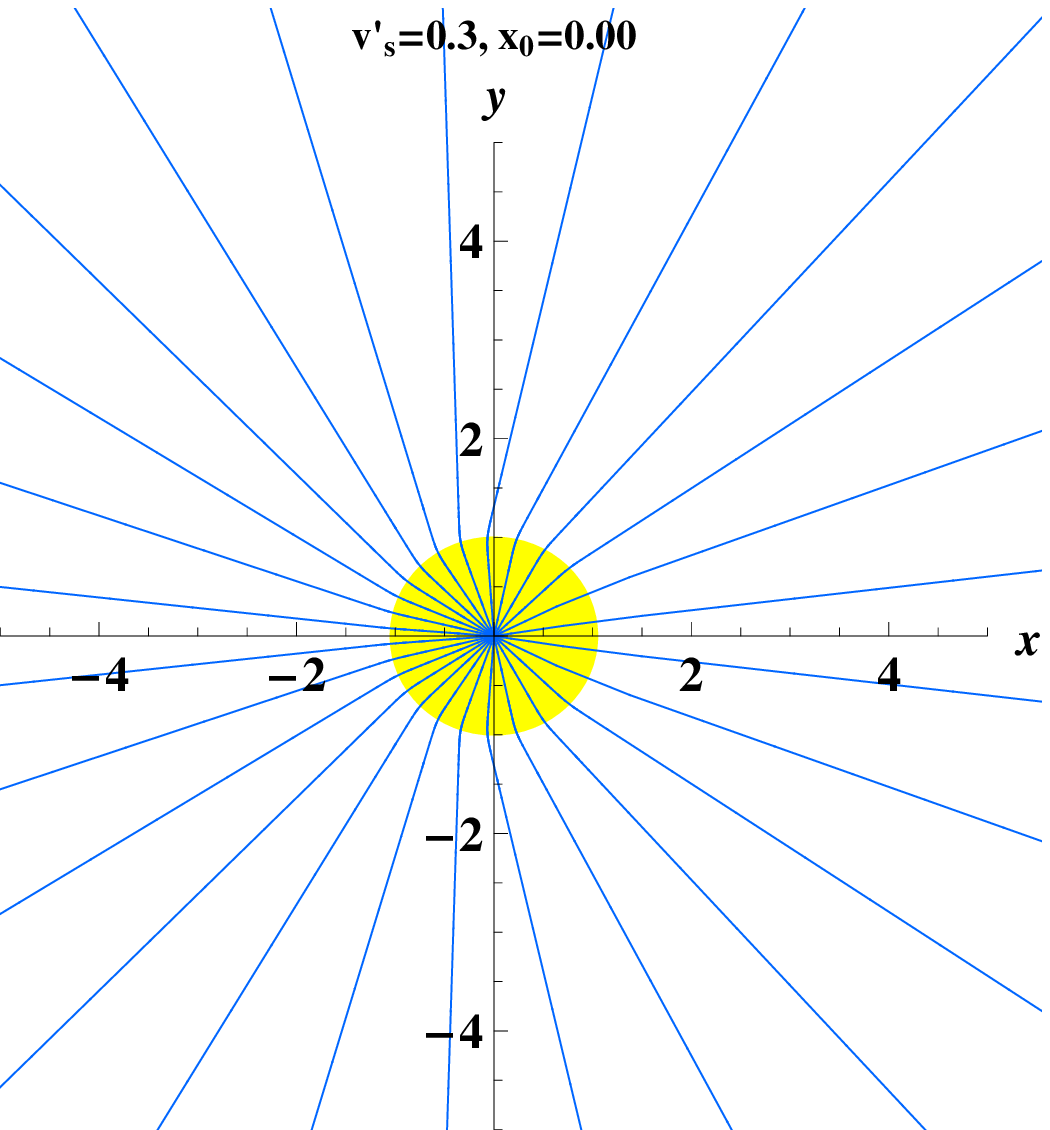,width=1.9in}
\epsfig{file=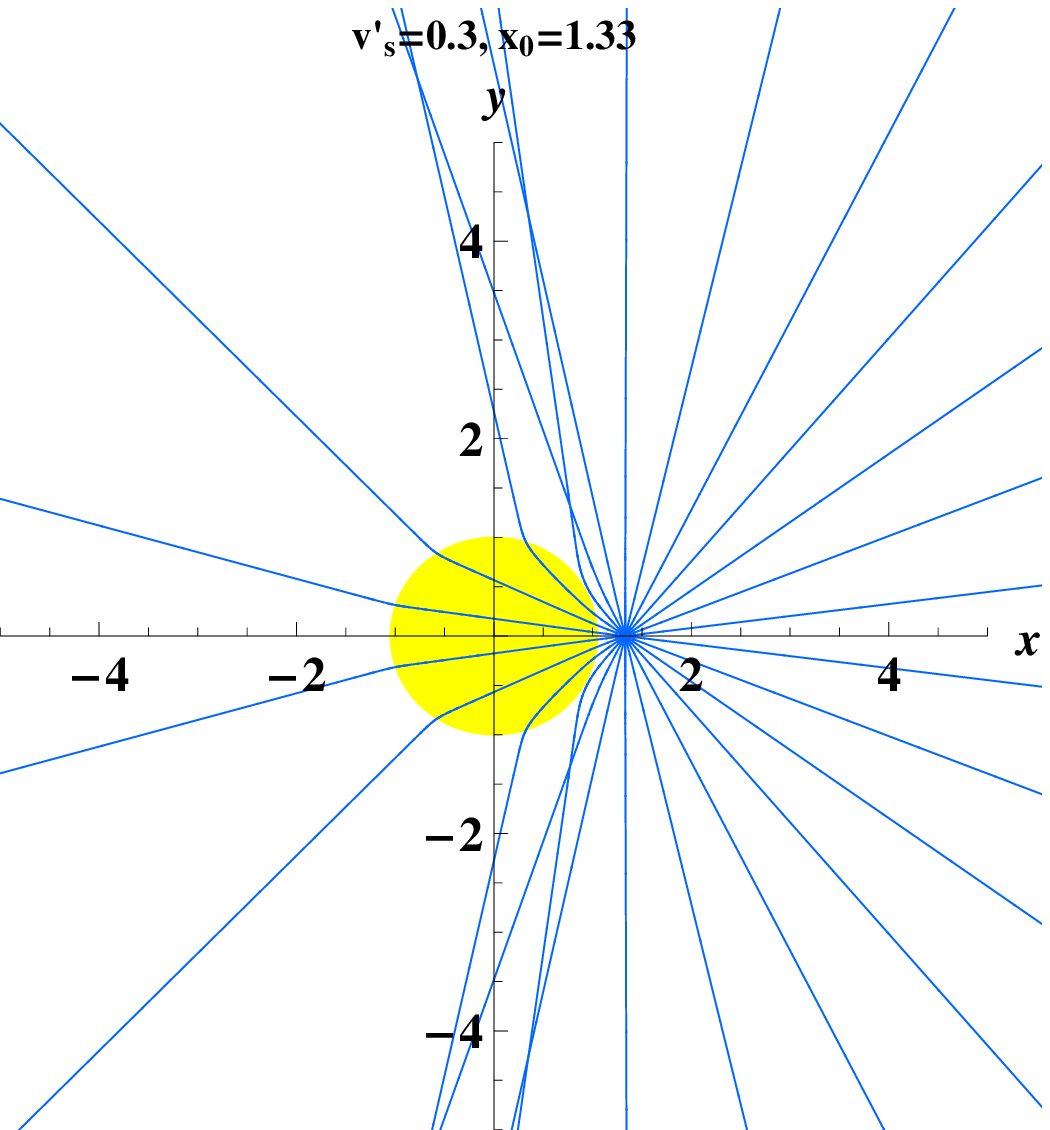,width=1.9in}\\
 \epsfig{file=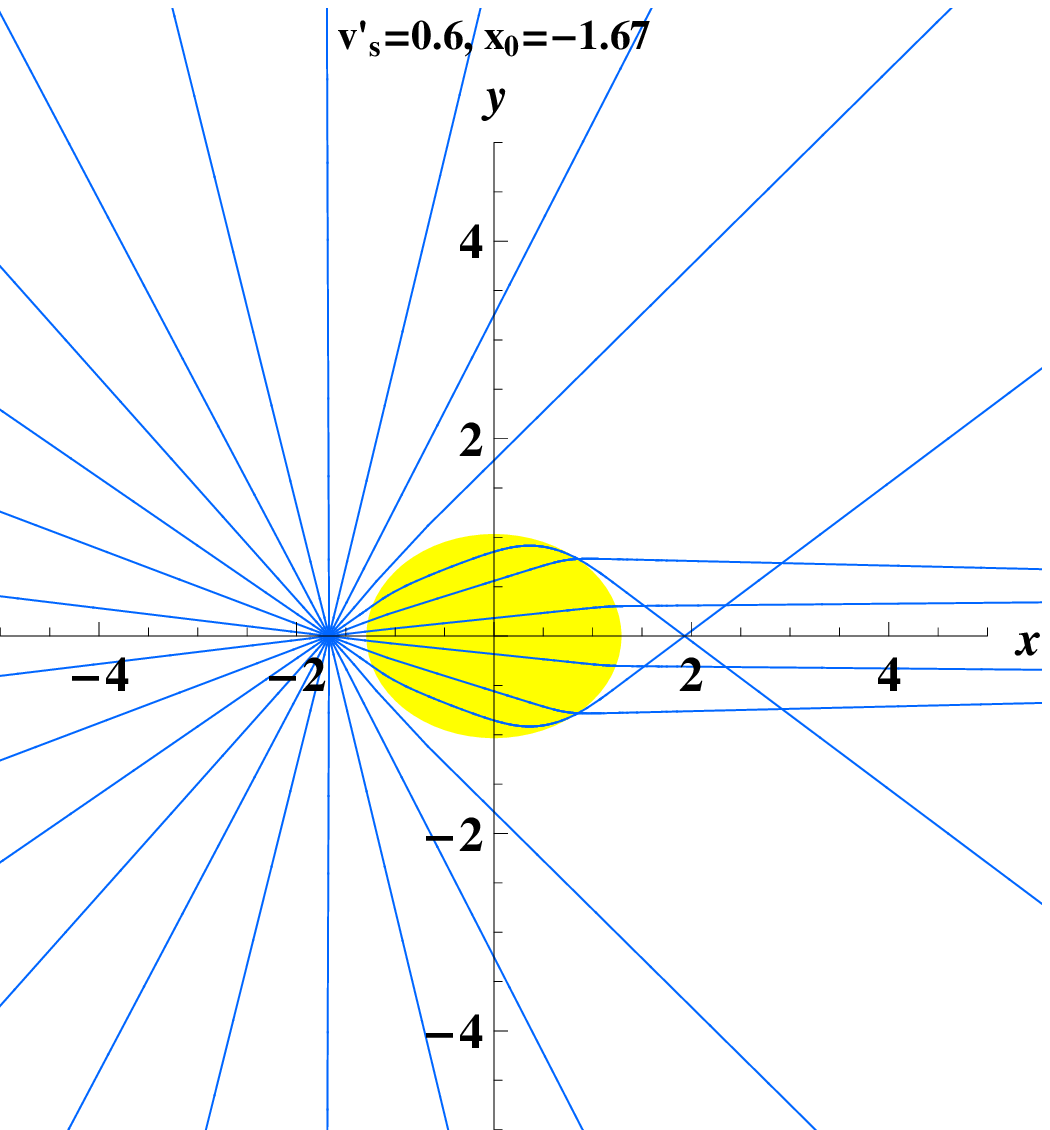,width=1.9in}
\epsfig{file=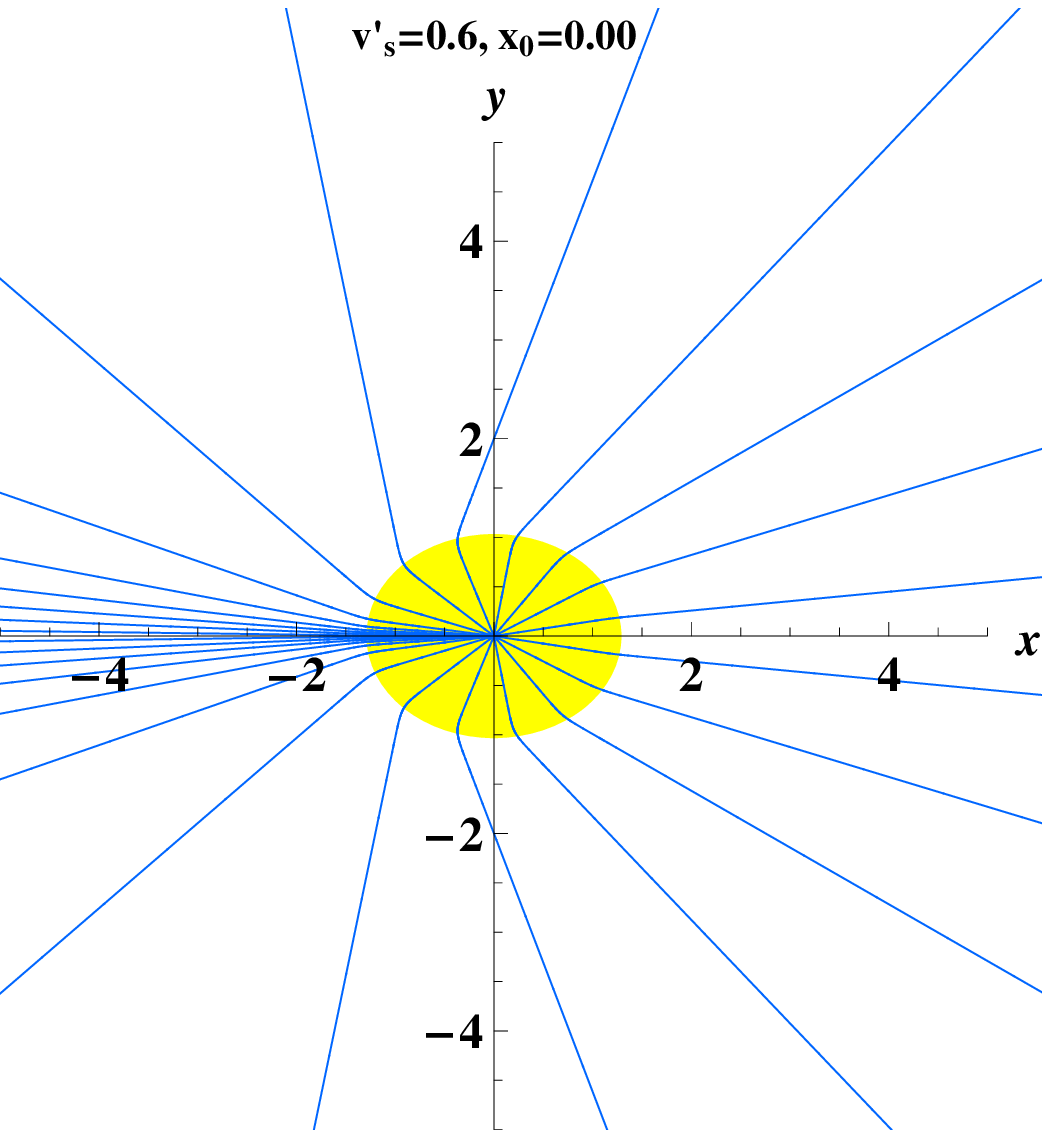,width=1.9in}
\epsfig{file=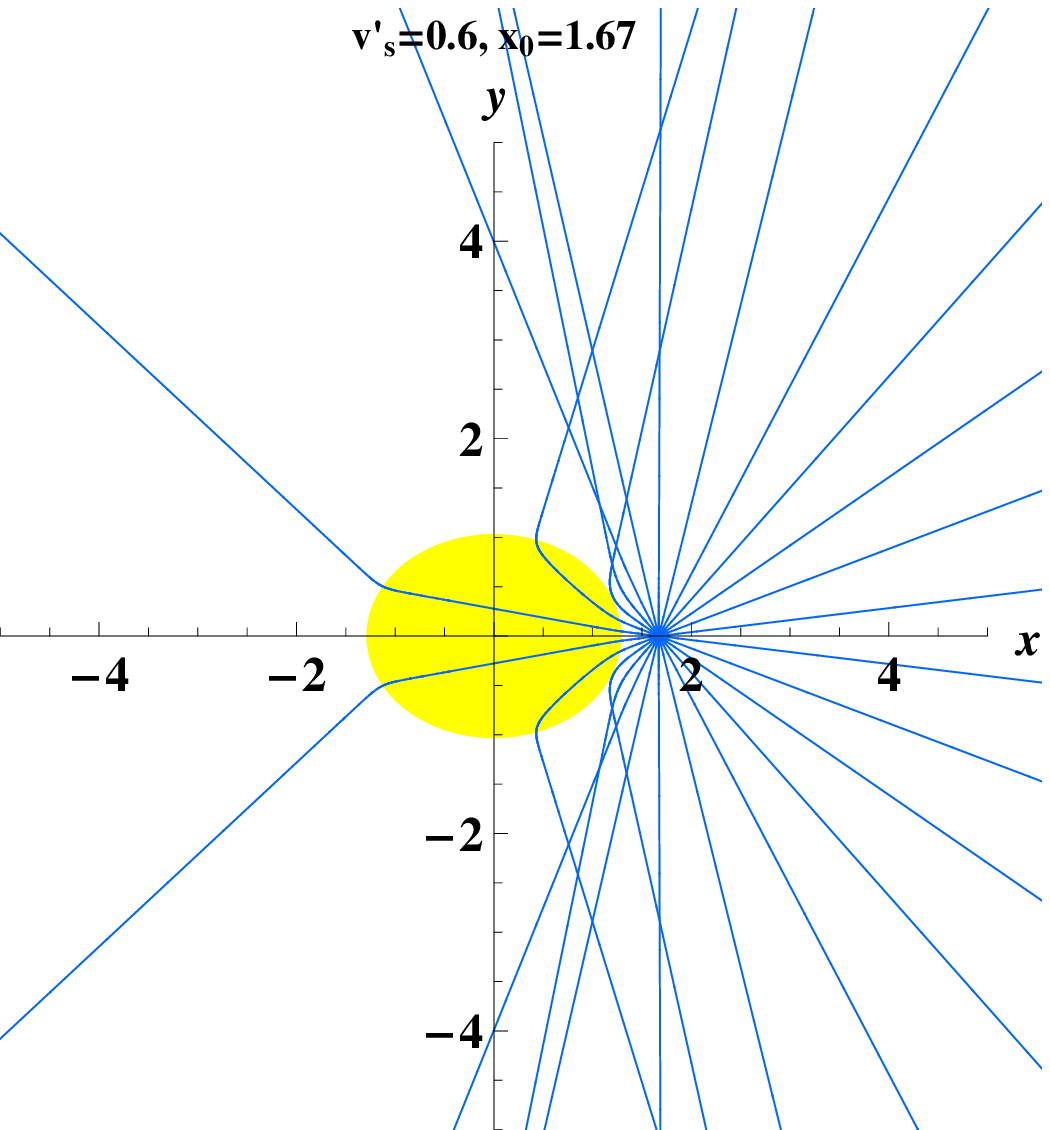,width=1.9in}\\
 \epsfig{file=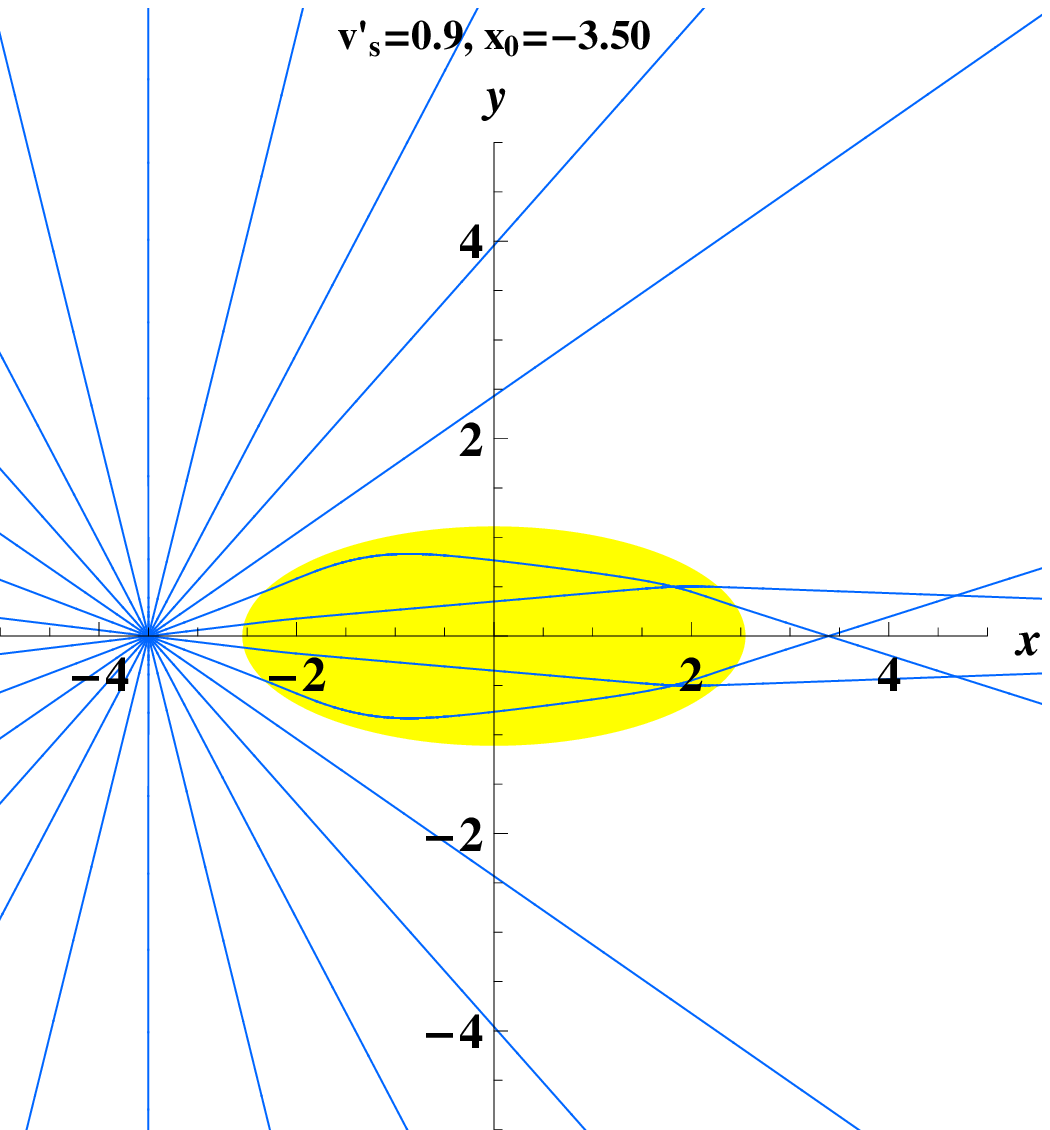,width=1.9in}
\epsfig{file=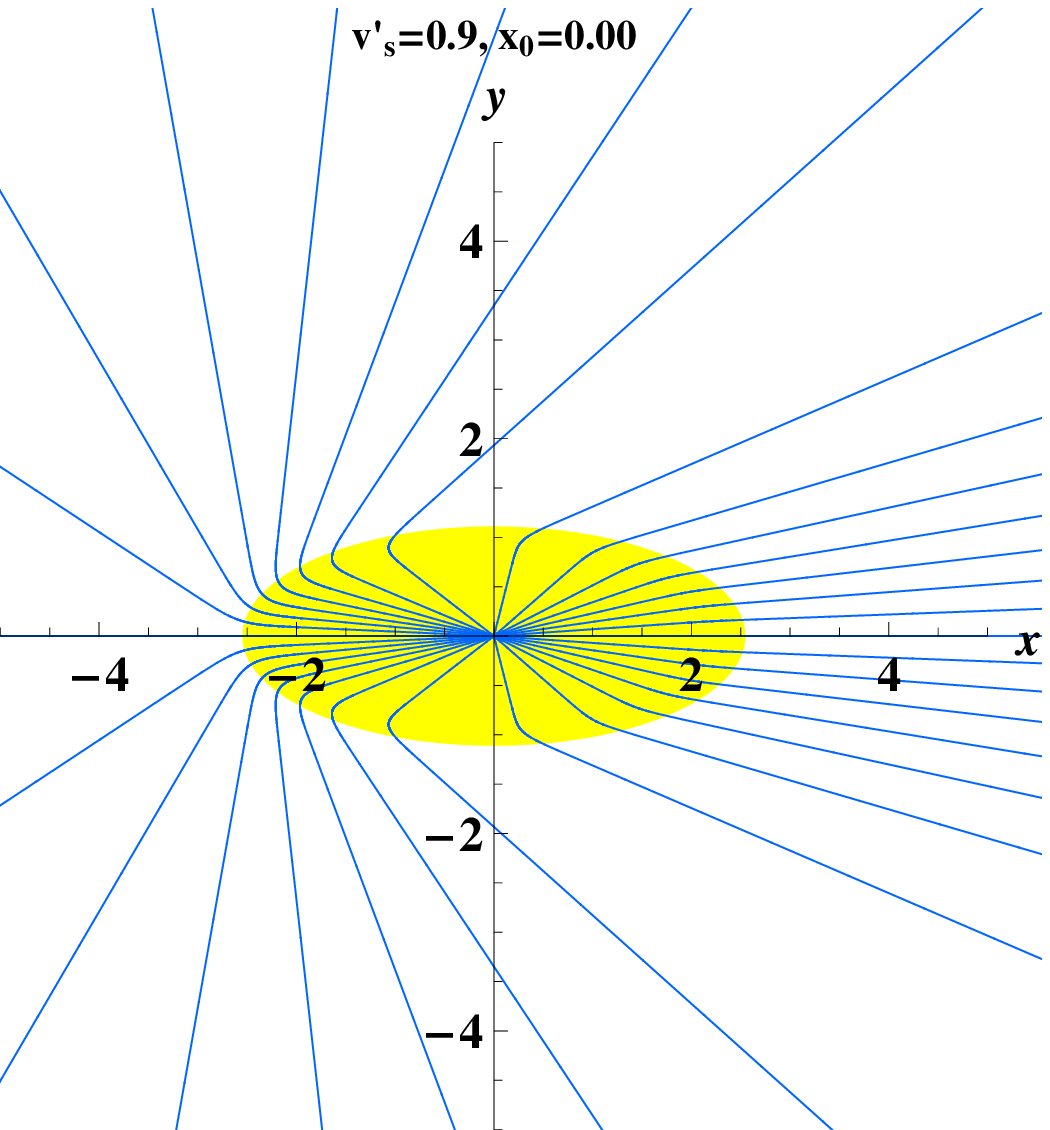,width=1.9in}
\epsfig{file=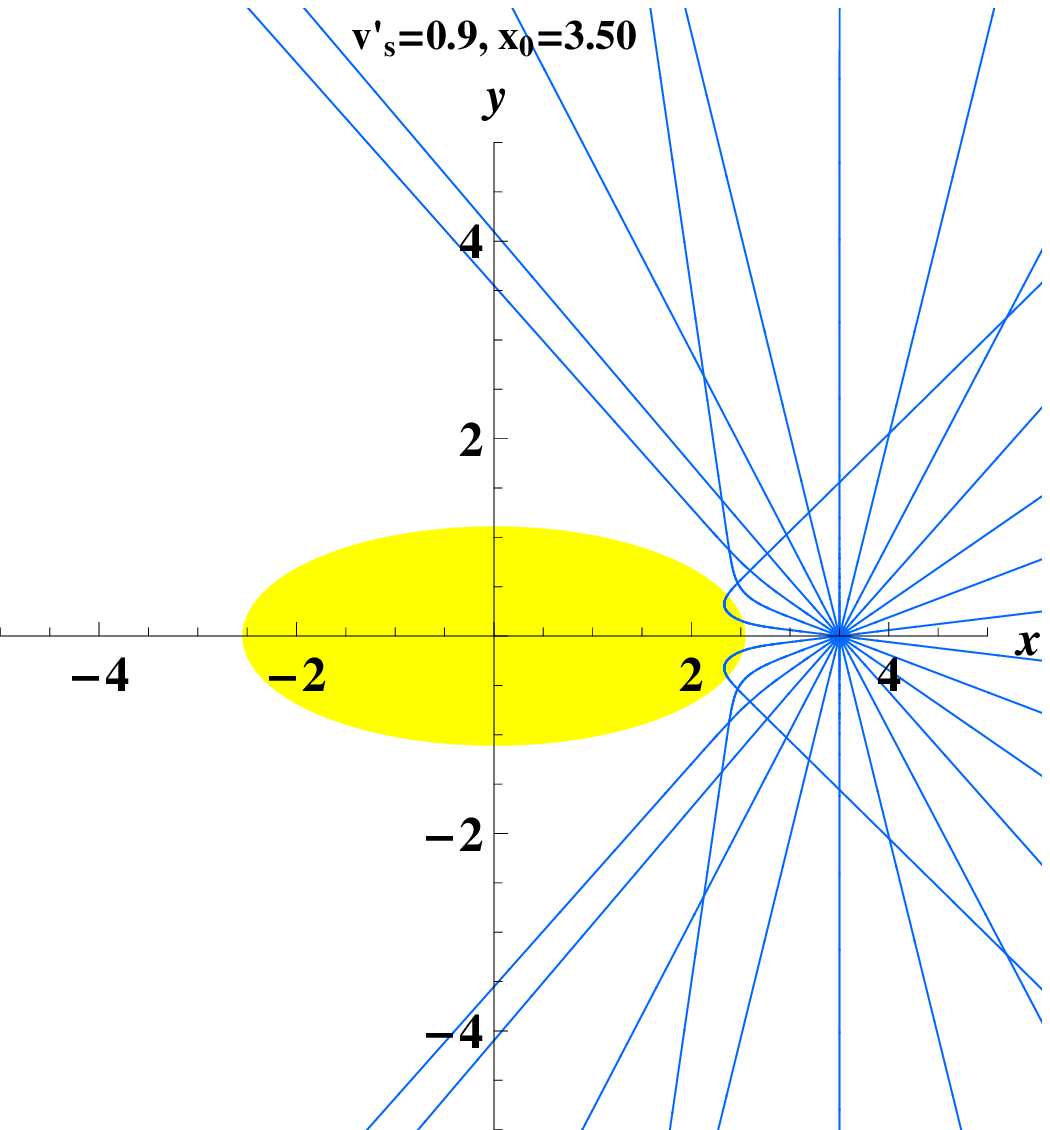,width=1.9in}
 \caption{Trajectories for rays in the $xy$ plane,
 emanating from sources on the $x$ axis, i.e.,   $\#r(0) =  x_0\,\hat{\#x}$,
 at equally spaced angular directions of $\#k(0)$, for $v'_s \in \lec 0.3,
0.6, 0.9 \ric$.
  Parameter values: $\sigma = 10$ and $R = 1$. }
\label{fig6}
\end{figure}

Rays initially propagating in radial directions in the $xy$ plane
 are represented in Fig.~\ref{fig6}
for $\sigma = 10$ and $R = 1$. We track rays which emanate from
point sources in the inner region (at the coordinate origin) and in
the outer region at  locations on the positive and negative $x$
axis; i.e., $\#r(0) =  x_0 \hat{\#x}$. The relative speed $v'_s \in
\lec 0.3, 0.6, 0.9 \ric$. Equally--spaced angular directions for the
initial relative wavevector $\#k (0)$ were considered. However, some
initial directions in the inner region correspond to evanescent
waves and these are not represented in Fig.~\ref{fig6}. The
proportion of directions which correspond to evanescent waves
increases as $v'_s$ increases for sources in the inner region.
Indeed, for $v'_s = 0.9$ with the source at the coordinate origin,
only $30\%$ of the possible $\#k (0)$ radial directions correspond
to propagating rays. The general trends apparent in
Figs.~\ref{fig3}--\ref{fig5} are also apparent in Fig.~\ref{fig6}.
That is, the inner region has a focusing effect for sources located
outside the inner region with $x_0 < 0$; for sources located outside
the inner region with $x_0
> 0$ ray trajectories tend to be progressively excluded from the
inner region as $v'_s$ increases.

\begin{figure}[!h]
\centering \psfull  \epsfig{file=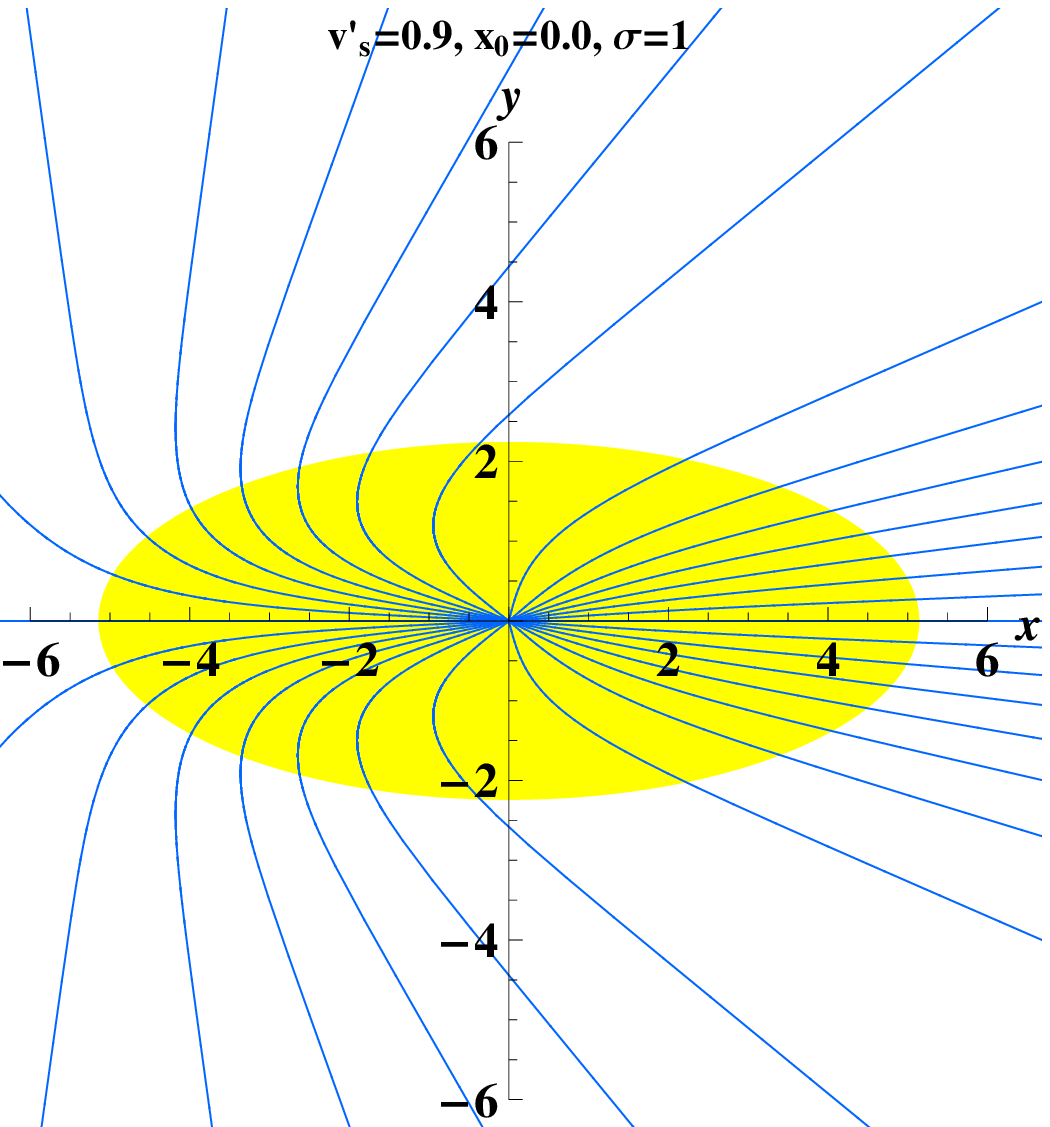,width=3.2in}
\epsfig{file=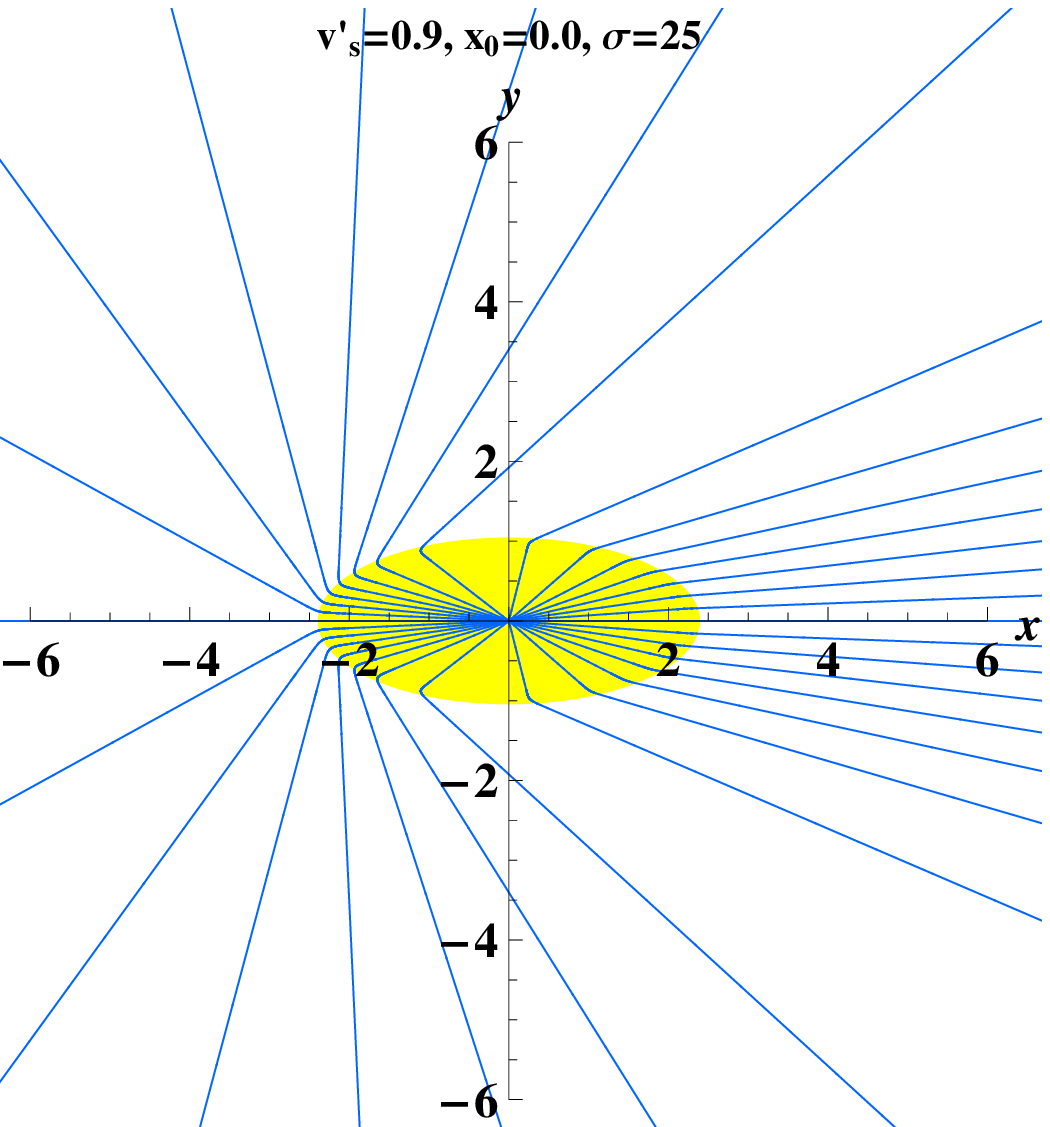,width=3.2in}
\caption{As
Fig.~\ref{fig6} except that $v'_s = 0.9$ and $\sigma
 \in \lec 1, 25 \ric$. } \label{fig7}
\end{figure}

Let us now turn to the influence of the thickness of the transition
zone between  the inner and outer regions, as dictated by the
parameter $\sigma$  via the scalar function $f$. In Fig.~\ref{fig7}
ray trajectories are provided which correspond to the scenario of
Fig.~\ref{fig6} but with $\sigma = 1$ and $25$. We note that $\sigma
= 25$ results in a more sharply defined top-hat profile with
straighter sides for $f$, whereas $\sigma = 1$ results in a profile
more rounded sides, as compared to $\sigma = 10$ which was used for
Fig.~\ref{fig6}. Comparing Figs.~\ref{fig6} and \ref{fig7}, we
deduce that, although the change in the direction of  rays at the
boundary between the inner and outer regions
  becomes more pronounced as $\sigma$ increases, the general
  pattern  of ray trajectories remains largely unaffected.

For clarity of representation,  ray trajectories restricted to the
$xy$ plane were considered in Figs.~\ref{fig3}--\ref{fig7}.
Trajectories of the same form can be observed in the $xz$ plane. The
trajectories for the $yz$ plane are likewise similar, albeit then
 the
 inner
 region of the constitutive-parameter space   is
 obviously circular in shape, regardless of the relative speed $v'_s$.

\section{Closing remarks}

A flat-spacetime representation of the Alcubierre spacetime has been
established by means of a Tamm medium which is asymptotically
identical to vacuum and has constitutive parameters
 which are  continuous functions of the
spatial coordinates.
 Thus,  the Tamm medium is amenable to  physical
realization as a nanostructured metamaterial. An alternative
approach~---~which utilizes a Galilean transformation instead of the
Lorentz transformation   \r{Lorentz}~---~gives rise to a Tamm medium
which is not asymptotically identical to vacuum and is accordingly
less well-suited to physical realization \cite{Smolyaninov_arxiv}.

Our geometric-optics study has revealed that ray trajectories are
acutely sensitive to the relative speed $v'_s$ of the warp bubble,
and  to the direction of its velocity $\#v'_s$, but
rather less sensitive to the thickness of the transition zone
between the  warp bubble and its background. In particular, for
rays which travel in the same direction as $\#v'_s$, the warp
bubble acts as a focusing lens, especially at large values of $v'_s$.

\vspace{10mm}

\noindent {\bf Acknowledgment:}  AL thanks the Charles Godfrey
Binder Endowment at Penn State for partial financial support of his
research activities.


\newpage

\newpage

\newpage

\newpage

\newpage

\newpage

\newpage

\end{document}